\documentclass[trackchanges,twocolumn]{aastex701}
\shorttitle{MIRI Counts Predictions Including AGN}

\begin{document}

\title{Modeling the {\sl JWST} MIRI Counts, Insights Into the Source Properties and Role of Dust-Obscured AGN}

\author[orcid=0009-0002-2209-4813,gname=Edgar,sname=Vidal]{Edgar Perez Vidal}
\affiliation{Department of Physics and Astronomy, Tufts University, Medford, MA 02155, USA}
\email[show]{edgar.vidal@tufts.edu}  

\author[orcid=0000-0002-1917-1200,gname=Anna, sname=Sajina]{Anna Sajina} 
\affiliation{Department of Physics and Astronomy, Tufts University, Medford, MA 02155, USA}
\email{Anna.Sajina@tufts.edu}

\author[gname=Amber,sname=Banks]{Amber Rose Banks}
\affiliation{Department of Physics and Astronomy, Tufts University, Medford, MA 02155, USA}
\email{amber.banks@tufts.edu}

\author[orcid=0000-0002-3915-2015, gname=Matthieu, sname=Béthermin]{Matthieu Béthermin}
\affiliation{Université de Strasbourg, CNRS, Observatoire Astronomique de
Strasbourg, UMR 7550, 67000 Strasbourg, France}
\email{matthieu.bethermin@astro.unistra.fr}

\author[orcid=0000-0001-6266-0213,gname=Carl, sname=Ferkinhoff]{Carl Ferkinhoff}
\affiliation{Department of Physics, Winona State University, Winona, MN 55987, USA}
\email{cferkinhoff@winona.edu}

\author[orcid=0000-0003-4030-3455, gname=Andrea, sname=Petric]{Andrea Petric}
\affiliation{William H. Miller III Department of Physics and Astronomy, Johns Hopkins University, Baltimore, MD 21218, USA}
\affiliation{Space Telescope Science Institute, 3700 San Martin Drive, Baltimore, MD 21218, USA}
\email{apetric4@jhu.edu}

\author[orcid=0000-0001-8592-2706, gname=Alexandra]{Alexandra Pope}
\affiliation{Department of Astronomy, University of Massachusetts, Amherst, MA 01003, USA}
\email{pope@astro.umass.edu}

\author[orcid=0000-0002-6221-1829,gname=Jianwei,sname=Lyu]{Jianwei Lyu}
\affiliation{Steward Observatory, University of Arizona, 933 North Cherry Avenue, Tucson, AZ 85721, USA}
\email{jianwei@arizona.edu}

\author[orcid=0000-0002-1912-0024, gname=Vivian, sname=U]{Vivian U}
\affiliation{IPAC, California Institute of Technology, 1200 E. California Blvd., Pasadena, CA 91125, USA}
\email{vivianu@caltech.edu}

\author[orcid=0000-0003-3466-035X,gname=L. Y,sname=Yung]{L. Y. Aaron Yung}
\affiliation{Space Telescope Science Institute, 3700 San Martin Drive, Baltimore, MD 21218, USA}
\email{yung@stsci.edu}

\author[orcid=0000-0002-9471-8499, gname=Pallavi, sname=Patil]{Pallavi Patil}
\affiliation{William H. Miller III Department of Physics and Astronomy, Johns Hopkins University, Baltimore, MD 21218, USA}
\email{ppatil13@jh.edu}






\begin{abstract}
Understanding the co-evolution of galaxies and active galactic nuclei (AGN) requires accurate modeling of dust-obscured systems. Recent surveys using the Mid Infrared Instrument (MIRI) onboard the  {\sl James Webb Space Telescope (JWST)} have uncovered a large population of dust obscured AGN, challenging current theoretical frameworks. We present an updated version of the Simulated Infrared Extragalactic Dusty Sky (SIDES) simulation framework. Our updates include modified star-forming and starburst galaxy spectral energy distribution (SED) templates as well as quiescent and AGN templates. We also incorporate a probabilistic assignment of the fraction of the IR emission that is due to an AGN. Our simulations successfully reproduce the observed MIRI source number counts, redshift distributions, and AGN population fractions. We find that AGN dominate at bright flux densities (\(S_\nu \gtrsim 20\, \mu\rm Jy\)) while main sequence galaxies dominate at the faint end. We also quantify the effects of cosmic variance, showing that surveys with areas below \(25\, \rm arcmin^2\) suffer from \(\sim 30 \% \) uncertainty in bright AGN counts. Finally, we provide diagnostic color-color diagrams and joint Near Infrared Camera (NIRCam) and MIRI flux distributions to aid interpretation of current and upcoming {\sl JWST} surveys.

\end{abstract}


\section{Introduction}\label{sec:Introduction}
A major goal of extragalactic astronomy is to understand the co-evolution of galaxies and AGN. However, much of the build-up of stars and black holes across cosmic time is heavily obscured by dust \citep[e.g.][]{Madau2014}. For dust obscured systems, the direct stellar photospheric or accretion disk emission is reprocessed by dust into the infrared part of the spectrum \citep[for a review see][]{lonsdale2006}. Often within the same system, both star-formation and AGN are present and contribute to the dust emission \citep[e.g.][]{Sajina2008, Petric2011,Petric2015, Kirkpatrick2015, Luo2025}. 

The mid-IR regime, roughly 3-30\,$\mu$m, is where the SEDs of purely star-forming galaxies, pure AGN and everything in between are most distinct, allowing us for example to find dust obscured AGN \citep[see][for recent reviews]{LacySajina2020,Lyu2022,Sajina2022}. Additionally, this region allows us to distinguish qunatitatively and consistently between the sources of IR emission even in complicated nearby objects like Luminous Infrared Galaxies, or highly obscured AGN hosts. At a population level, this regime was first opened up with the deep {\sl ISO} ISOCAM 15\,$\mu$m surveys \citep{Elbaz1999,Gruppioni2002} and later with {\sl Spitzer} IRAC and MIPS 24\,$\mu$m surveys \citep[e.g.][]{Marleau2004,Papovich2004,Ashby2009}. Phenomenological models were developed to help interpret the {\sl ISO} and later {\sl Spitzer} data \citep[see e.g.][and references therein]{lagache2004,Pearson2005,Rowan-Robinson2009}. They revealed a rapidly evolving dust-obscured population and relatively low, but increasing with increasing flux, contributions of AGN to the counts even at 15\,$\mu$m consistent with observational constraints \citep{Fadda2002}. However, later studies highlighted the need to account for dust obscured AGN within AGN-starburst composites, which significantly increase the role of AGN within the mid-IR population \citep[e.g.][]{Kirkpatrick2015}  

With the advent of the {\sl James Webb Space Telescope} \citep[{\sl JWST};][]{Gardner2006, Gardner2023} with its Mid-IR instrument \citep[MIRI;][]{Rieke2015}, we now have both significantly better spatial resolution and greater sensitivity, orders of magnitude improvement of obscured AGN identifications than {\sl ISO} or {\sl Spitzer} surveys. Moreover, with 9 broad filters between 5 and 25\,$\mu$m, MIRI allows a much finer sampling of the mid-IR spectral energy distribution (SED) than earlier instruments, which is particularly important for the selection of dust obscured AGN \citep[][]{Rieke2024,Lyu2024}. In the last couple of years, several MIRI imaging surveys have published their counts \citep[e.g.][]{Yang2023,Sajkov2024,Stone2024,Harish2025}. Using multi-filter MIRI data, a large population of dust obscured AGN are revealed \citep{Yang2023, Lyu2024}, and we now have observed MIRI AGN counts as well \citep{Lyu2024}. 

On the theoretical side, the newly published MIRI counts have been modeled using both empirical \citep{Bisigello2021,Rowan-Robinson2024,Rowan-Robinson2025} and semi-analytic \citep{Cowley2018,Lagos2019,Manzoni2025} models. These models show broad agreement with the observations, but also reveal some key problems. For example, some semi-analytic models exhibit an overprediction at lower fluxes and an underprediction at higher fluxes \citep[e.g.][]{Yang2023}. This discrepancy is larger at longer wavelengths, suggesting issues with the dust SEDs. At shorter wavelengths, while providing a reasonable overall fit to the 5.6\,$\mu$m counts, the \citet{Cowley2018} model predicts quenched galaxies dominate in this regime, whereas observations suggest a dominance of main-sequence intermediate-redshift galaxies \citep{Sajkov2024}. Lastly, these models have a limited treatment of AGN populations, especially AGN-starburst composites. This limited treatment of AGN populations highlights the need to extend galaxy formation models to fully incorporate the properties of MIRI-discovered dust-obscured AGN. The mid-IR selected dust obscured AGN discovered by \citet{Lyu2024}, for example, are not only more numerous than previous studies had suggested but also do not show the drop-off in obscured fraction with increasing AGN luminosity that was previously seen using {\sl Spitzer} surveys \citep{Lacy2015}.

In this paper, we address the above modeling issues by building upon the framework of the Simulated Infrared Dusty Extragalactic Sky (pySIDES) code\footnote{\url{https://gitlab.lam.fr/mbethermin/sides-public-release}}\citep{Bethermin2017}. This code has already been demonstrated to reproduce the counts pre-JWST counts, especially from mid-IR out to the far-IR. pySIDES not only models the population statistics, but also simulates the infrared sky and is therefore invaluable in planning for future surveys \citep{Bethermin2024}. Here we introduce three key improvements to the core pySIDES model, which are critical to modeling the newly discovered {\sl JWST}-MIRI galaxies and AGN, making it the most promising modeling framework for this work. These include: 1) the star-forming and starburst galaxies templates now include the stellar contribution as well as a quiescent galaxy population; 2) an AGN SED library including starburst-AGN composites; and 3) a probabilistic prescription for the likely contribution of an AGN to the IR luminosity of any of our modeled galaxies. We assess the effects of our modified SED library vs the underlying population assumptions by presenting model predictions both within the core pySIDES framework as well as using the Santa Cruz SAM lightcone \citep[SC SAM; see][and references therein]{Yung2023} but with the same IR SED library. This effectively means updating the dust emission modeling of the SC SAM lightcone leading to more robust joint modeling of the source populations in existing and upcoming NIRCam+MIRI surveys.

The primary goal of this work is to improve pySIDES predictions for {\sl JWST}-MIRI sources while also providing a modeling framework useful for planning for and interpreting data from future far-IR missions such as the PRobe far-Infrared Mission for Astrophysics \citep[PRIMA; e.g. see][]{Donnellan2024, Bethermin2024}. Our paper is organized as follows. In Section~\ref{sec:Simulatons}, we provide an overview of the pySIDES and SC SAM galaxy formation models. Section~\ref{sec:Data_Methods} describes our methodology including SED library modifications and approach to modeling the AGN population. In Section~\ref{sec:Results}, we present our results including the comparison of modeled and observed number counts across the 8 different MIRI filters; population breakdown; redshift distributions per flux level; and color diagnostics. We provide a discussion and potential application for current and future programs in Section~\ref{sec:discussion}. In Section~\ref{sec:Summary}, we present our summary and conclusions. Throughout this paper, we adopt the cosmological parameters consistent with 
\citet{PlanckIII2016}: \(h = 0.678\), \(\sigma_8 = 0.823\), 
\(\Omega_\Lambda =  0.693\), \(\Omega_\mathrm{M} = 0.307\), and \(\Omega_\mathrm{b} = 0.048\), \(n_s = 0.96\).

\section{Simulations}\label{sec:Simulatons}
\subsection{PySIDES}\label{subsec:SIDES semi-empircal}

For our study, we make use of the pySIDES semi-empirical code. The core model thereof is detailed in \citet{Bethermin2012,Bethermin2017}, but below we summarize the key ingredients.

As the starting point, pySIDES utilizes the Bolshoi-Planck dark matter-only Cosmological simulation \citep{Rodriguez-Puebla_2016}. The total volume of the simulation is \((250 \, h^{-1} \mathrm{Mpc})^3\), with a dark-matter particle resolution \(1.5 \times 10^8 \,h^{-1} M_\odot\). Dark matter halos and their properties are identified with \textsc{Rockstar} and \textsc{Consistent Trees} \citep{Behroozi2013, Behroozi2013B}. For each halo and subhalo, a stellar mass is assigned based on abundance matching. Typically, a monotonically increasing stellar mass with some scatter is assigned to an increasing halo mass. Instead, pySIDES assigns a stellar mass based on the maximum circular velocity \(v_{\mathrm{pk}}\), which has been shown to be more closely correlated with stellar mass \citep[e.g.][]{Reddick2013, Gkogkou2023}. The assigned mass is drawn to reproduce the observed stellar mass function (SMF). The SMF is assumed to be a smooth double Schechter \citep{Baldry2012} whose parameters are constrained to match local, intermediate redshifts \(z<4\) and high redshift galaxies \(z>4\) \citep[e.g][]{Kelvin2014, Grazian2015, Moutard2016, Davidzon2017}. 

pySIDES includes SED libraries for two populations: main sequence star-forming galaxies and starbursts.  A probabilistic approach is taken where first galaxies are assigned as either star-forming or quiescent based on their stellar mass and redshift and known constraints on the quiescent fraction. Once identified, quiescent galaxies are not considered further in pySIDES as they do not contribute to the longer wavelength data that were of primary interest. The star-forming galaxies are assigned a star-formation rate (SFR) following the 2SFM formalism of \citet{Sargent2012,Bethermin2012}. Additionally, the evolution of the fraction of starburst galaxies is probabilistically assigned following \citet{Sargent2012, Hopkins2010}. At \(z \le 1\), the fraction of star-forming galaxies classified as starbursts increases monotonically with redshift from \(\sim 1.5 \% \) to \(\sim 3 \%\). At \(z \ge 1\), this fraction is fixed at \(3\% \). 

The SEDs for the main sequence and starburst galaxies use the templates from \citet{Magdis2012}. These SED templates vary with the mean intensity of the radiation field \( \langle U \rangle\), a parameter that strongly correlates with the dust temperature and evolves with redshift. The \( \langle U \rangle\) used for a given galaxy depends on its type (main sequence or starburst) and redshift alone, following the observationally constrained parametric form with intrinsic scatter \citep[][]{Magdis2012,Bethermin2015, Bethermin2017}. All SEDs in pySIDES are normalized to $L_{\rm IR, tot}= 1\,L_\odot$. Each template is then scaled by the galaxy’s \(L_{\rm IR, tot}\), derived from its SFR using the \citet{Kennicutt1998} conversion, to produce the observed fluxes. Throughout this paper, $L_{\rm IR, tot}$ is defined as the integral of the IR SED from \(8-1000 \,\mu\)m. Finally, magnification \(\mu\) is applied for strongly and weakly lensed sources using the prescriptions of \citet{Hezaveh2011} and \citet{Hilbert2007}, respectively. 

With all of the above elements, pySIDES generates a 2 \(\rm deg^2\) lightcone. This lightcone includes both fundamental parameters such as sky positions, redshifts, stellar masses, SFRs, M${_{\rm halo}}$, and $L_{\rm{IR}}$ but also observed flux density in a range of user-specified infrared filters for all galaxies. pySIDES allows for a user-supplied lightcone, which bypasses their core population model and only utilizes the following quantities: stellar mass, halo mass, redshift, SFR, and classification of quiescent, main sequence star-forming, or starburst galaxies. We make use of this feature when presenting SC SAM results throughout this paper. 

\subsection{The Santa Cruz Semi-Analytic Models}\label{subsec:SC_SAM}
This work also utilizes the 2-deg$^2$ lightcones presented in \citet{Yung2023}, each containing galaxies over the redshift range $0 < z \lesssim 10$ and rest-frame magnitudes $-22 \lesssim M_{\rm UV} \lesssim -16$ that are simulated with the Santa Cruz semi-analytic model \citep[SC SAM;][]{Somerville1999, Somerville2015}. These lightcones are constructed from dark matter halos extracted from the Small MultiDark-Planck (SMDPL) simulation in the MultiDark suite \citep{Klypin2016}. For each halo, a Monte Carlo merger tree is generated on the fly using the extended Press-Schechter formalism \citep{Somerville1999a}, within which the SC SAM predicts a wide range of physical and observable properties. We refer the reader to \citet{Yung2022} and \citet{Somerville2021} for further details regarding the construction of these lightcones.
 
The SC SAM incorporates a suite of carefully calibrated physical processes to track galaxy evolution, including cosmological accretion, gas cooling, star formation, chemical enrichment, and stellar and AGN feedback. The simulated galaxy populations have been shown to reproduce observed trends across a wide range of redshifts, including rest-frame UV luminosity functions up to $z \sim 10$ \citep{Yung2019}, stellar mass functions up to $z \sim 8$ \citep{Somerville2015, Yung2019}, and two-point auto-correlation functions up to $z \sim 7.5$ \citep{Yung2022}. The SC SAM also includes prescriptions for the growth of supermassive black holes through mergers and accretion \citep{Somerville2008}, and the predicted AGN populations have been shown to reproduce the observed AGN bolometric luminosity functions between $2 \lesssim z \lesssim 6$ \citep{Yung2021}.

Specifically, we use realization 0 and apply a redshift cut \(z<6\) of the Ultra-Wide\footnote{\url{https://flathub.flatironinstitute.org/group/sam-forecasts/sam-ultra-wide}}\citep{Yung2023, Somerville2021} 2 deg\(^2\) simulation. Model galaxies here have SEDs which are based on the \citet{Bruzual_and_Charlot2003} stellar population synthesis. The rest-frame SEDs were used to calculate monochromatic luminosities, including observed magnitudes in several optical and near-infrared filters.

\section{Methodology}\label{sec:Data_Methods}

The primary goal of the original SIDES implementation was to reproduce the far-infrared (FIR) number counts. In this framework, quiescent galaxies are assumed not to emit in the FIR and were therefore assigned a SFR = 0, which sets \(L_{\rm IR} = 0\). Additionally, it does not include AGN, as their contributions to the FIR emission are negligible \citep[e.g.][]{Bethermin2012, Bethermin2017}, although in extreme and rare cases this may not be the case \citep{Hickox2018, Mckinney2021}. However, we need to include both Quiescent and AGN here as they contribute to the mid-infrared populations as discussed in the Introduction. We make use of the SC SAM lightcone to inform the incorporation of both quiescent galaxies and AGN into pySIDES. In addition, we test the effect of different population assumptions by feeding the SC SAM lightcone itself into pySIDES as described above. Below, we begin with a brief outline of how we accomplish this step. 

\subsection{Implementing the SC SAM Lightcone Into pySIDES}\label{subsec:implenting_SC_SAM}

As discussed above, in order to test out different model population assumptions, pySIDES allows for a user-supplied lightcone, which bypasses every step until the IR SEDs are assigned and flux density calculations are done.  This lightcone requires SFR and star-forming vs. quiescent classification for each model galaxy. 

\begin{figure}
    \centering
    \includegraphics[width=\columnwidth]{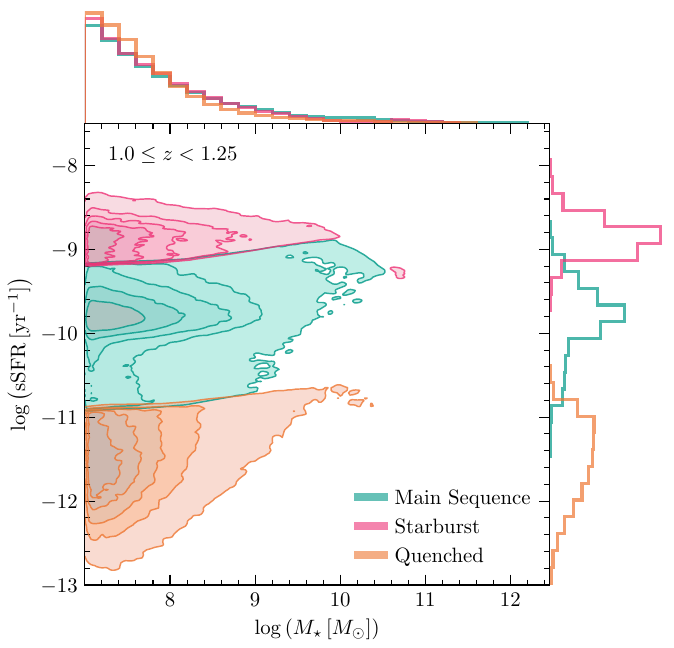}
    \caption{Kernel density estimation (KDE) contours of galaxies from the SC SAM, color by galaxy classification (Main Sequence, Starburst, Quenched) based on their position in the sSFR vs \(M_\star\) plane. Marginal histograms of sSFR and \(M_\star\) are shown along the top and right axes, respectively. See Section~\ref{subsec:implenting_SC_SAM} for classification details.}
    \label{fig:sSFR_vs_Mstar}
\end{figure}

We classify SC SAM galaxies as quiescent, main sequence star-forming or starburst using the specific star formation rate\(\text{ sSFR = SFR}/M_\star\) vs \(M_\star\) relation in various redshift bins. For each redshift slice, we define quiescent galaxies as those lying 1 dex below the median sSFR in bins of stellar mass, corresponding to galaxies that produce stars at a rate 10 times lower than the typical population. Starburst galaxies are defined to have sSFRs 5.3 times higher than the median. An example is shown in Figure \ref{fig:sSFR_vs_Mstar}. 

For all star-forming galaxies, the infrared luminosity coming from star formation alone, \(L_\mathrm{IR, SF}\), is calculated from the SFR assuming the \citet{Kennicutt1998} conversion relation. This is needed as all pySIDES star-forming galaxy SEDs are normalized to \(L_\mathrm{IR, tot}=1 \, L_\odot\) and need to be multiplied by the true \(L_\mathrm{IR, tot}\) of each galaxy. Therefore, the above steps were sufficient to feed this lightcone into pySIDES and generate expected IR flux densities for all model galaxies using the standard SED library. 

\subsection{Quiescent Galaxies} \label{subsec:implenting_quiescents}
\begin{figure}
    \centering
    \includegraphics[width=\columnwidth]{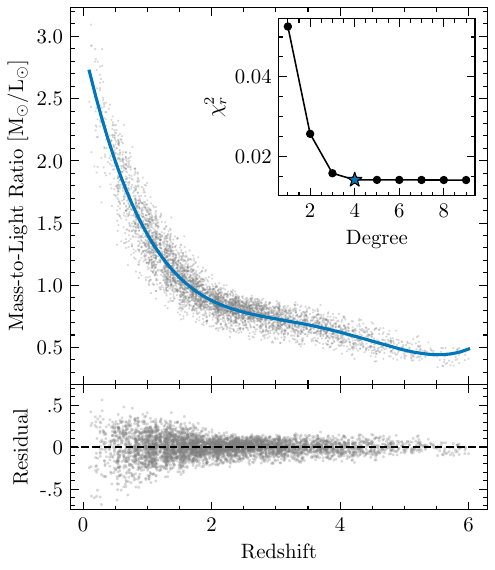}
    \caption{Mass-to-light ratio at the 1.22 \(\mu\)m \(J_{\rm rest}\) band for Quiescent galaxies as a function of redshift, based on the SC SAM. A polynomial function of degree 6 fit is shown in blue, and the residuals are shown in the lower panel. For a given redshift, each quiescent galaxy is assigned a mass-to-light ratio using this functional form to de-normalize the Ell2 template.}
    \label{fig:mass_to_light_vs_redshift}
\end{figure}
Infrared dust emission from quiescent galaxies is ignored in pySIDES because at a given mass, the contribution to the FIR is an order of magnitude less than that of star-forming galaxies \citep[e.g.][]{Viero2013, Amblard2014, Man2016, Gobat2017, Gobat2018}. However, at redshifts $\gtrsim$ 2.5, the shortest wavelength MIRI filters (e.g. F560W and F770W) start to sample the 1.6\,$\mu$m stellar bump. Therefore, quiescent galaxies can make a non-negligible contribution to the counts. Quiescent galaxies are already flagged in pySIDES, and we flag them in the SC SAM lightcone as described in Section~\ref{subsec:implenting_SC_SAM}. However, there is no SED assigned to them. We adopt the Ell2 template from the SWIRE template library \citep{Polletta2007}, which is a template for an elliptical galaxy with a stellar population age of 2\,Gyr. In practice, our results in Section~\ref{sec:Results} are insensitive to the specific choice of elliptical template (e.g. Ell5), as the SED is scaled using the rest frame $J$-band mass-to-light ratios (see below).

While the star-forming SEDs are normalized by $L_{\rm{IR, tot}}$, which depends on the SFR, this normalization cannot be applied to quiescent galaxies. We chose instead to normalize the quiescent galaxy SED templates to the \(J_{\rm rest}\) band magnitudes as shown in Figure~\ref{fig:SEDs_MS_SB_Ell2}. For the SC SAM lightcone, de-normalization can be performed directly using the tabulated $J_{\rm rest}$-band magnitudes. However, these are not available when running the standard pySIDES. Instead, we compute the stellar mass-to-light ratio in bins of redshift for all quiescent galaxies in the SC SAM and de-normalize the Ell2 SED using stellar mass and the expected \(J_{\rm rest}\) band mass-to-light ratio at a given redshift.

Figure~\ref{fig:mass_to_light_vs_redshift} shows the \(J_{\rm rest}\)-band mass-to-light ratio as a function of redshift of all galaxies flagged as quiescent in the SC SAM lightcone. This trend is expected due to the age of the stellar population \citep[e.g. see][]{2001Bell_de_Jong}. We tested rest-frame near-infrared bands (\(J,\, H,\, \text{and}\, K\)) but adopted the  \(J_{\rm rest}\) mass-to-light ratio due to its minimal scatter. A fourth-degree polynomial is fit to the data in Figure~\ref{fig:mass_to_light_vs_redshift}. To avoid overfitting, we perform a cross-validation test, evaluating the reduced chi-squared on a held-out test set and selecting the polynomial order where the test chi-squared stops decreasing (see inset in Figure~\ref{fig:mass_to_light_vs_redshift}). We additionally verified the polynomial order selection using the Akaike information criterion (AIC) and Bayesian information criterion (BIC), computed assuming Gaussian errors. The AIC decreases sharply from degree 1 to degree 4, from $-1.97 \times 10^5$ to $-2.70 \times 10^6$, with negligible further improvement beyond degree 4 ($<0.1\%$ decrease from degree 4 to 9). The BIC yields effectively identical values, as the log-likelihood improvement from degree 1 to degree 4 far exceeds the additional per-parameter penalty of BIC relative to AIC. Both criteria confirm a fourth-degree polynomial as the optimal model, consistent with the cross-validation result. This best-fit model allows us to de-normalize the quiescent galaxy SED in pySIDES using only its stellar mass and redshift.

\subsection{Modified Star-Forming Galaxies Spectral Energy Distributions}\label{subsec:sed assignment}

\begin{figure}
    \centering
    \includegraphics[width=\columnwidth]{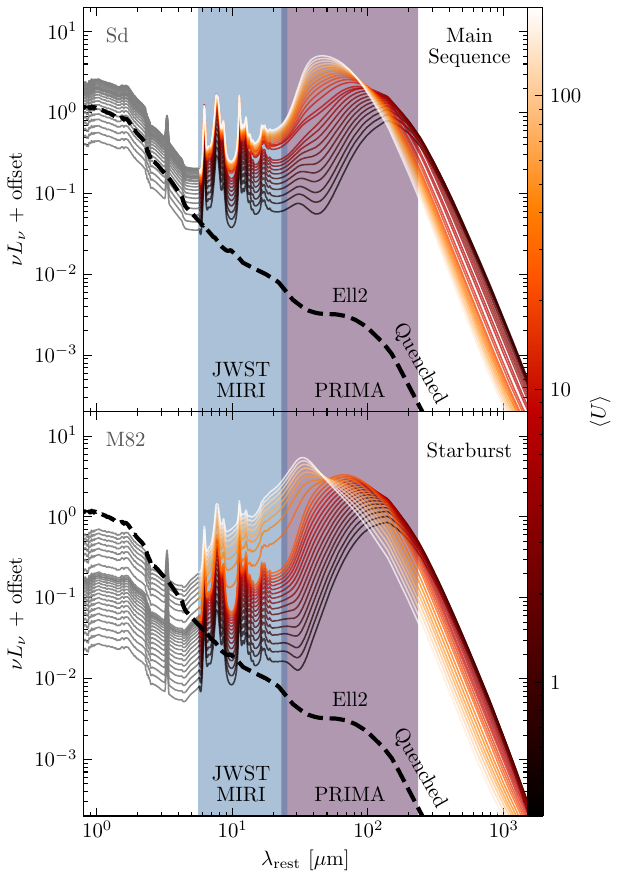}
    \caption{A sample of the modified Main Sequence and Starburst Galaxy SEDs from \citet{Magdis2012} normalized to \(L_{\rm IR, tot} = 1\,L_\odot\). The color bar indicates the mean intensity of the radiation field \(\langle U \rangle\), and the SEDs are offset for clarity. Stellar contributions for the Main Sequence and Starburst galaxies are modeled by the \emph{Sd} and \emph{M82} templates (grayscale), respectively, while quenched galaxies are represented by the Ell2 template (black-dashed, normalized to \(J_{\rm rest}\)); all from the SWIRE library \citep{Polletta2007}. The stellar contributions are stitched for visualization purposes and do not reflect scaling by stellar mass. The {\sl JWST} MIRI and PRIMA imager coverages are indicated by blue and purple shaded regions, respectively. Details on SED modifications are provided in section~\ref{subsec:sed assignment}.}
    \label{fig:SEDs_MS_SB_Ell2}
\end{figure}

Because of redshifting, the shorter wavelength MIRI filters sample stellar photospheric emission rather than dust emission. Therefore, we modify the core (dust-only) main sequence and starburst SED libraries to include stellar emission. Note that stellar emission does not vary significantly in the restframe $\gtrsim$ 1.6 \(\mu\)m regime, which is mostly what we sample with our redshifted MIRI filters. Therefore, it is beyond the scope of this project to do detailed modeling of the star-formation histories of individual galaxies. Thus, we adopt reasonable single stellar emission templates for main sequence and starburst galaxies, respectively, using the \emph{Sd} and \emph{M82} templates from the SWIRE library \citep{Polletta2007}. The total SED for each star-forming or starburst galaxy is constructed by summing the stellar emission, scaled by stellar mass, and the dust component, scaled by IR luminosity. Thus, by construction, our modified SEDs account for the evolving star-forming galaxy main sequence.

In the treatment of the star-forming galaxies stellar emission, we follow the same procedure used for quiescent galaxies (Section~\ref{subsec:implenting_quiescents}), beginning with computing the mass-to-light ratio as a function of redshift for galaxies flagged as main sequence or starbursts. This approach allows us to scale the stellar emission by mass and accurately reproduce the total stellar contribution across galaxy populations. Example stellar+dust emission SEDs are shown in Figure \ref{fig:SEDs_MS_SB_Ell2} where the color corresponds to \(\langle U \rangle\) and the stellar contributions are shown in gray scale. However, note again that individual galaxies will have SEDs where the stellar and dust components are scaled to correspond to each galaxy's stellar mass and star formation rate, respectively. 

\subsection{AGN SED Templates}\label{subsec:AGN_SEDs}
\begin{figure}
    \centering
    \includegraphics[width=\columnwidth]{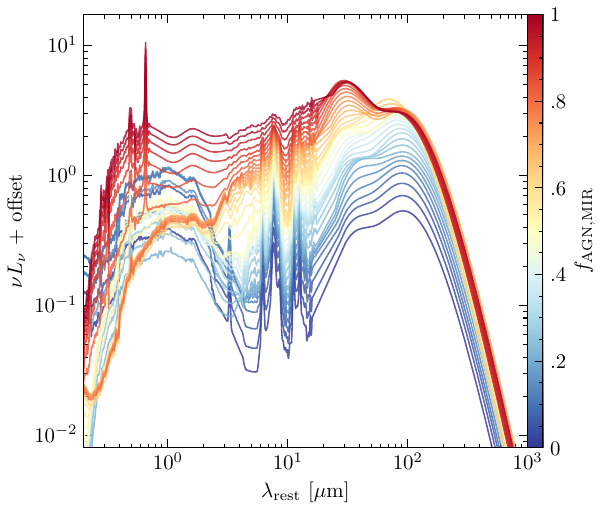}
    \caption{Sample of the modified MID IR SED template catalog from \citet{Kirkpatrick2015} normalized to \(L_{\rm IR, tot} = 1 \, L_\odot\).  The SEDs are offset for easier interpretation.  We adapt the definition of star forming galaxies as \(f_{\rm AGN, MIR} < 0.2\), composite galaxies  \(f_{\rm AGN, MIR} \in [0.2-0.8]\), and AGN as \(f_{\rm AGN, MIR} > 0.8\). The procedure for adding the stellar contribution to the SED is summarized in Section~\ref{subsec:AGN_SEDs}. Note that only modeled galaxies with \(f_{\rm AGN} > 0.2\) will adopt this template library}
    \label{fig:AGN_SED}
\end{figure}

We model the IR emission of AGN using the SED template library from \citet{Kirkpatrick2015}. This library is based on a sample of 300+ cosmic noon ($z\sim1-3$) sources with {\sl Spitzer}-IRS mid-IR spectra as well as {\sl Herschel} far-IR photometry. The sources are all classified from the AGN fraction ($f_{\rm AGN,MIR}$) of the mid-IR light. This fraction is calculated from an empirical model fit to the IRS spectra, where the model represents a measure of the strength of the PAH features relative to the continuum.  The sources are all then binned by $f_{\rm AGN,MIR}$ in steps of 0.1. The IRS spectra are averaged within each bin and a two temperature model is fit to the binned far-IR data. The result is 11 SED templates spanning the full range of 0-1 in $f_{\rm AGN,MIR}$.  We adopt these templates as they are strongly rooted in observations of galaxies in the critical cosmic noon regime expected to dominate our MIRI counts \citep[see e.g.][]{Sajkov2024}.  

As with the SEDs of star-forming galaxies, we need to account for the stellar bump within these templates. We cannot adopt the same Sd/M82 SWIRE templates we used before, as they would not be appropriate for stronger AGN templates with higher \(f_{\rm AGN}\), which include additional near-IR and mid-IR emission coming from the AGN; therefore we examined the SWIRE template library (which again is very empirically driven) and decided upon the best template to use for the stitching based on the overall optical-to-IR SED being consistent with real galaxies within the SWIRE template library. For composites with \(f_{\rm AGN,MIR} \in [0.2-0.5]\), we stitched NCG 6240 at $2.1\, \mu m$. For composites \(f_{\rm AGN,MIR} \in [0.6,0.8]\) we stitched I19254-S at $3 \, \mu m$. Lastly, for the AGN-dominated templates (\(f_{\rm AGN,MIR} > 0.8\)), we stitched the QSO2 template at $3 \, \mu m$.  We chose these points as they would not affect the \(L_{\rm IR, tot}\) normalization. For $f_{\rm AGN, MIR} < 0.2$, our simulations use the updated \citet{Magdis2012} SED templates with the Sd  template stitched at \(5.6 \mu m\) as shown in Figure~\ref{fig:SEDs_MS_SB_Ell2}. 

For consistency with the pySIDES main sequence and starburst templates, we convert all templates to $\nu L_{\nu}$ and normalize by $L_{\rm{IR}}$. Finally, to create a more finely sampled SED library, we use a weighted sum of two nearest templates. To generate the target SED \(T_{\rm target}\) with arbitrary \(f_{\rm AGN,MIR,target}\), let \(T^-\) and \(T^{+}\) denote the SEDs corresponding to adjacent \(f_{\rm AGN,MIR}\) SEDs in template grid such that \(T^{-} \le T_{\rm target} \le T^{+}\). The \(T_{\rm target}\) is generated using linear interpolation. 

\begin{equation}\label{eq:AGN_fine_grid}
    T_{\rm target} = \left(1-w\right) T^- + w T^{+}\\
\end{equation}

\begin{equation}\label{eq:AGN_fine_grid_weights}
    w = \frac{T_{\rm target} - T^{-}}{T^{+} - T^{-}}
\end{equation}

where \(w\) is the weight that ensures the sum of the templates preserves the normalization. Using this method, we increase our library to 2001 SED templates with \(\Delta f_{\rm AGN,MIR} = 0.0005\). The final AGN SED template library we adopt is given in Figure~\ref{fig:AGN_SED}, where the different templates are colored by $f_{\rm AGN,MIR}$. While all SEDs are normalized by $L_{\rm{IR}}$ here, they are offset for clarity of presentation. 

Following \citet{Kirkpatrick2015}, we defined galaxies with \(f_{\rm AGN,MIR} < 0.2\) as star-forming, those with \(f_{\rm AGN,MIR} \in [0.2-0.8]\) as composites, and those with \(f_{\rm AGN,MIR} > 0.8\) as AGN. Note that galaxies with \(f_{\rm AGN,MIR} < 0.2\) will automatically use the star-forming (SF) galaxies SEDs shown in Figure~\ref{fig:SEDs_MS_SB_Ell2} and only those classified as composites or AGN adopt the SEDs in Figure~\ref{fig:AGN_SED}. However, we verified that the \(f_{\rm AGN,MIR} = 0.0\) template here is consistent with Main Sequence star-forming galaxy template with $\langle U \rangle$ appropriate for a cosmic noon galaxy.

\subsection{Assigning \texorpdfstring{$f_{\rm{AGN}}$}{f\_AGN} to Model Galaxies}\label{subsec:determining_f_agn}

\begin{figure}
    \centering
    \includegraphics[width=\columnwidth]{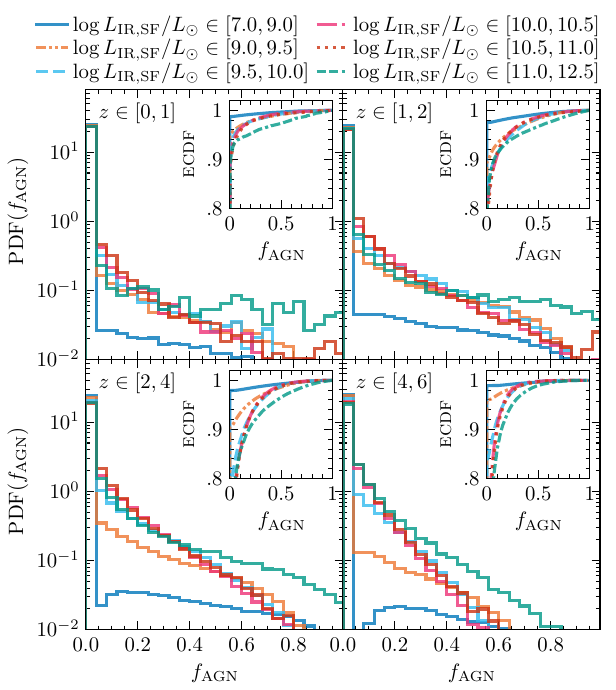}
    \caption{The normalized probability density functions (PDF) of \(f_{\rm AGN}\) in bins of redshift and \(L_{\rm IR,SF}\) using \(\alpha = 1\) (see text for details). The inset plots are the ECDF used for the inverse sampling procedure as described in Section \ref{subsec:determining_f_agn}. }
    \label{fig:f_agn_lum_and_redshift_distributions}
\end{figure}

Our AGN SED templates are defined by their $f_{\rm AGN,MIR}$ as described above. This can be converted to the total IR AGN fraction \(f_{\rm AGN, tot} \) (hereafter \(f_{\rm AGN}\) for simplicity), using the quadratic relation given in \citet{Kirkpatrick2015}. To assign the appropriate AGN SED template to a given galaxy, we first determine the probability of a given $f_{\rm AGN}$, which is expected to vary as a function of redshift and IR luminosity. 

We use the SC SAM lightcone to produce probability density functions (PDFs) for $f_{\rm AGN}$ based on each galaxys SFR and BHAR, which are then converted to $f_{\rm AGN}$. The AGN contribution to the total infrared is defined as 

\begin{equation}\label{eq:fraction_of_agn}
    f_{\rm AGN} = \frac{L_{\rm IR,AGN}}{L_{\rm IR,tot}}
\end{equation}

where \(L_{\rm IR,AGN}\) is the integrated 8-1000\,$\mu$m AGN luminosity and \(L_{\rm IR,tot}\) is the integrated \(8-1000 \, \mu \rm m\) luminosity including contributions from both stars and AGN. 

Therefore, we have $L_{\rm IR,tot}=L_{\rm IR,AGN}+L_{\rm IR,SF}$ where $L_{\rm IR,SF}$ is the IR luminosity due to star-formation alone.  We convert the SC SAM SFRs averaged over the past 100\,Myr to \(L_{\rm IR,SF}\) using the \citet{Kennicutt1998} conversion for consistency with pySIDES.

The AGN accretion disk luminosity is related to the BHAR ($\dot M$) by

\begin{equation}\label{eq:Black_Hole_Acretion_rate}
L_{\rm disk} = \epsilon  \dot M c^2 
\end{equation}

where \(\epsilon\) is the radiative efficiency typically assumed to be 0.11 \citep[e.g. see][]{2012Hirschmann, Lacy2015}. We note that \(\epsilon \) is known to have a large scatter \citep[e.g. see][]{Davis&Laor2011}. For this work, we fixed \(\epsilon \) to a single value in our model to remain consistent with the assumption adopted in hydrodynamical simulations such as \citep[Horizon-AGN;][]{Dubois2014};\citep[EAGLE;][]{Schaye2015}; \citep[SIMBA;][]{Dave2019}. However, below we discuss a test we performed in varying $\epsilon$ over a range of values spanning from radiatively inefficient to radiatively efficient ($\epsilon$ = 0.05-0.3). 
The AGN luminosity in the IR is assumed to be related to the disk luminosity as:

\begin{equation}\label{eq:disk_and_agn_luminosity}
    L_{\rm IR,AGN}= \alpha L_{\rm disk}
\end{equation}
where \(\alpha\) denotes the fraction of the luminosity of the accretion disk that is reprocessed in the IR. The parameter \(\alpha\) is expected to vary with values close to 1 for the most heavily obscured systems and roughly 0.3-0.5 for Type-1 AGN \citep{Lyu2017}. 

We use three different assumptions about $\alpha$ ($\alpha=0.3, \,0.5\, \text{and } 1.0$) to compute \(f_{\rm AGN}\) for all sources in the SC SAM. When testing the SC SAM directly, we use the tabulated values of BHAR to compute deterministic \(f_{\rm AGN}\) values for each galaxy using Equations~\ref{eq:fraction_of_agn}--\ref{eq:disk_and_agn_luminosity}. In contrast, within the pySIDES framework, we draw \(f_{\rm AGN}\) probabilistically from the normalized probability density functions (PDF) of \(f_{\rm AGN}\) in bins of \(L_{\rm IR, SF}\) and redshift. These are shown in Figure~\ref{fig:f_agn_lum_and_redshift_distributions} in the case of $\alpha=1$. 

These PDFs are then sampled via the inverse transformation method, which is done by constructing the empirical cumulative distribution function (ECDF), as opposed to the binned cumulative distribution function, which is sensitive to binning. The ECDF is defined as, 

\begin{equation}
    \label{eq:emeprical_cumulative_distribution_function}
    \hat{F}(x) = \frac{1}{N} \sum_{i=1}^{N} \mathbf{1}_{x_i \leq x}
\end{equation}

\begin{figure*}
    \centering
    \includegraphics[width=\textwidth]{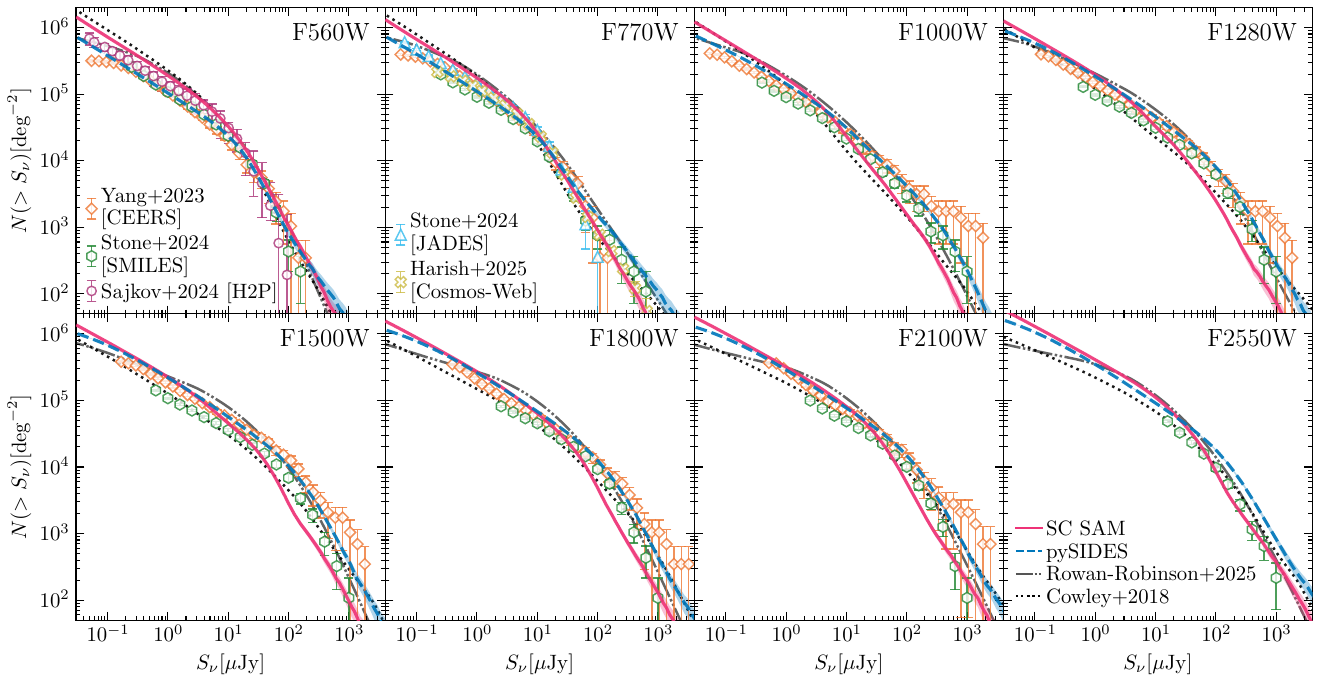}
    \caption{Integral number counts across 8 MIRI imaging filters assuming \(\alpha = 0.55\). The data points come from several different surveys as labeled (see Section~\ref{subsec:number_counts} for details). We overlay the {\sc evolfb} model from \citet{Cowley2018} as well the ones from \citet{Rowan-Robinson2025}. Our modified pySIDES and SC SAM models are shown in dashed blue and solid red lines, respectively. The shaded region in our models represents the Poisson statistical error.}
    \label{fig:miri_counts}
\end{figure*}

\begin{figure*}
    \centering
    \includegraphics[width=\textwidth]{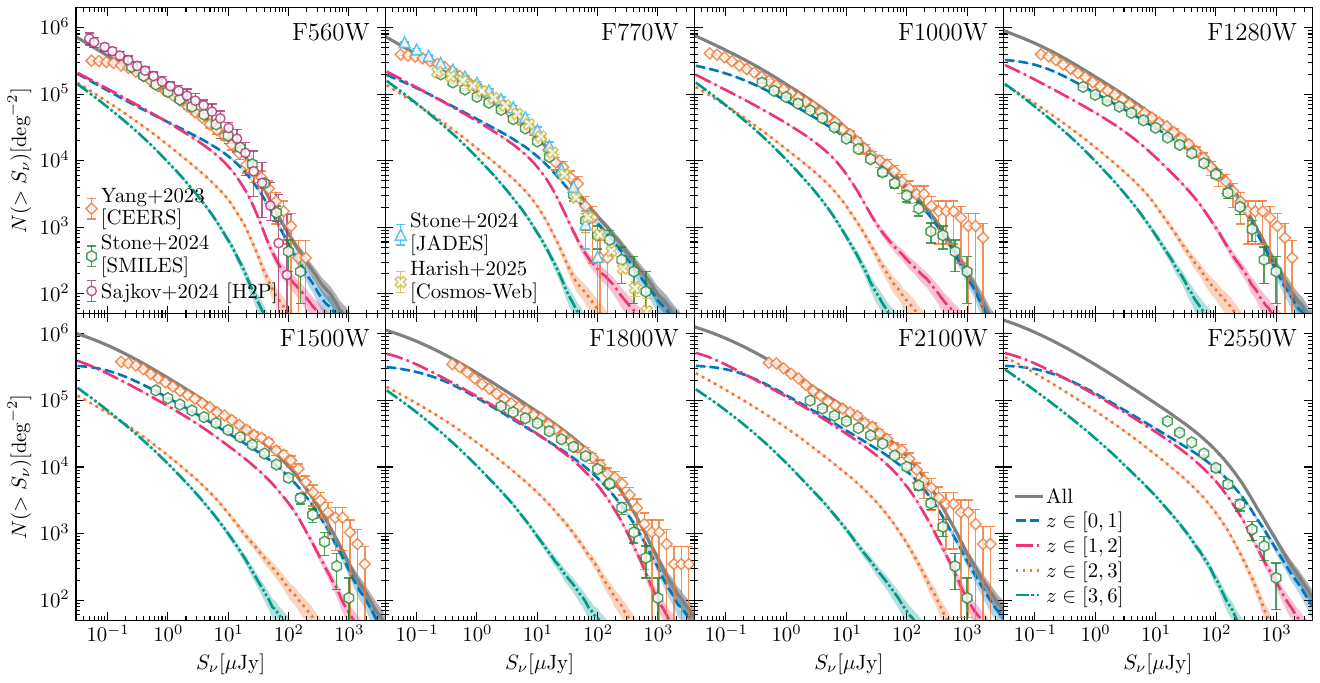}
    \caption{Integral number counts of our modified pySIDES model assuming \(\alpha = 0.55\), separated by cumulative redshift contribution. Each curve represents the integral counts in bins of redshift. The data points from Figure~\ref{fig:miri_counts} are shown for comparison. The shaded regions show the Poisson statistical uncertainty.}
    \label{fig:miri_counts_redshift_split}
\end{figure*}
where \( \mathbf{1}_{x_i \leq x} \) denotes the indicator function, equal to 1 if \( x_i \leq x \), and 0 otherwise. To sample \(f_{\rm AGN}\) within a given redshift and \(L_{\rm IR, SF}\) bin, we generate a random number \(x \sim \rm U \, [0,1]\) and apply the inverse transform. Large samples are generated to verify that the PDF is accurately recovered. 

The binned PDF and corresponding ECDF are shown in Figure \ref{fig:f_agn_lum_and_redshift_distributions} for bins of \(\log L_{\rm IR, SF}/L_\odot\, \in [10^7-10^{12.5}]\) and \(z \in \, [0,6]\).  We find that the binning choice only affects the results at low redshifts, where most of the evolution occurs. Therefore, we use bins \(z \in [0,1]\) and \(z \in [1,2]\) for \(z \le 2\) and broader bins for \(z > 2\) to ensure sufficient samples sizes and smooth ECDFs. For sources with values outside the defined bins, the nearest bin is used to assign an \(f_{\rm AGN}\) value. 

We also tested the effect of varying $\epsilon$. Specifically,  we tested a range of values spanning from radiatively inefficient to radiatively efficient ($\epsilon$ = 0.05-0.3). 
As expected, the effect of higher $\epsilon$ values is a shift toward higher $f_{\rm AGN}$ values in the distribution shown in Figure~\ref{fig:f_agn_lum_and_redshift_distributions} which increases the total number of AGN and vice versa for lower values of $\epsilon$. We find that the total predicted MIRI number counts (discussed in Section~\ref{subsec:number_counts}) are unaffected however, the number of AGN (i.e. the AGN counts alone, discussed in \ref{subsec:counts_by_population}) would change. In Section\,\ref{subsec:population_fractions} we also explore the effect of varying $\alpha$ on the AGN counts. Because of their coupled nature, the $\alpha$ value that best matches the observed AGN counts does depend on our adopted choice of $\epsilon$. We stress that our adopted values for $\epsilon$ or $\alpha$ are not intended to represent a unique solution, but a choices that constrains the observed MIRI AGN counts from \citet{Lyu2024}.

These distributions allow us to assign $f_{\rm{AGN}}$ values probabilistically to any model galaxy given its $L_{\rm{IR, SF}}$ and redshift. Since our AGN SED library is in terms of $f_{\rm{AGN,MIR}}$, we invert the quadratic relation from \citet{Kirkpatrick2015} to convert $f_{\rm{AGN}}$ to $f_{\rm{AGN,MIR}}$. Sources with \(f_{\rm AGN, MIR} < 0.2\) (and not flagged as quiescent) will default to the updated main sequence or starburst templates. Sources classified as composite \((f_{\rm AGN, MIR} \in [0.2,0.8]\)) or AGN (\(f_{\rm AGN, MIR} > 0.8\)) will use the updated MIR library. To scale the SEDs, we compute $L_{\rm{IR,tot}}$ from each galaxies $L_{\rm{IR,SF}}$ (derived from the SFR) and $f_{\rm AGN}$ via

\begin{equation}
    \label{eq:total_luminosity_with_AGN}
    L_{\rm IR,tot} = \frac{L_{\rm IR,SF}}{1 - f_{\rm AGN}}
\end{equation}

 To avoid unreasonable values of \(L_{\rm IR,tot}\), including divide by zero errors, we impose a maximum of  \(f_{\rm AGN}\)= 0.9.

\section{Results} \label{sec:Results}
\subsection{Number Counts}\label{subsec:number_counts}

The modified pySIDES as well as SC SAM+pySIDES models produce a \(2\,\rm deg^2\) lightcone where each object has a flux density measurement in each MIRI filter. From these lightcones, we compute the integral number counts \(N(>S_\nu)\) by directly counting the number of galaxies with flux density greater than a given threshold \(S_\nu\) and dividing by the total area \(A_{\rm eff} = 2\, \rm deg^2\).

Figure~\ref{fig:miri_counts} compares the observed MIRI counts vs. our modified pySIDES, SC SAM, and the earlier GALFORM model of \citet{Cowley2018}. We show 8 of the 9 MIRI imaging filters (F1130W is omitted due to the dearth of observations with it). We compare these models with observations from multiple MIRI imaging surveys, including: Halfway to the Peak \citep[H2P;][]{Sajkov2024}; Cosmic Evolution Early Release Science Survey \citep[CEERS;][]{Yang2023}; \citep[SMILES, JADES;][]{Stone2024}; and COSMOS-Web \citep{Harish2025}. To ensure consistency with prior work, we verify that our implementation preserves the FIR number counts established in \citet{Bethermin2017}, as shown in Figure~\ref{fig:herschel_counts}. This validation confirms that our updates do not compromise agreement at longer wavelengths.

Our model broadly reproduces the observed number counts across all 8 MIRI filters. The modified pySIDES model is generally closer to the counts than the SC SAM or the \citet{Cowley2018} models. At the shorter wavelengths (F560W and F770W), both the SC SAM and the \citet{Cowley2018} models overpredict the counts at the faint end. At these wavelengths, the pySIDES performs better at the faint end. Due to redshifting, these bands are more sensitive to rest-frame optical emission rather than dust emission. Since we use the same SED template libraries, the discrepancy likely reflects inherent galaxy population differences. We explore this possibility in more detail in Section~\ref{subsec:population_differences_modeling}. 

Both models tend to overpredict the counts at the faint end (e.g see F1000W panel). This may indicate issues with the modeling or underestimated incompleteness in the observational data. For instance, in \citet{Sajkov2024}, the incompleteness is estimated based on the recovery of fake injected sources onto images where real sources are masked. However, considering the source density, some degree of blending is expected, which would boost the calculated completeness factors, increasing the observed counts at the faint end. 

The \citet{Cowley2018} semi-analytic model with its two feedback prescriptions (baseline and EvolFB), also broadly captures the shape of the observed counts. But it tends to also overpredict at the faint end and underpredict at the bright end. Additionally, as discussed further in Section~\ref{subsec:population_fractions}, the predicted nature of the MIRI population differ from our updated models as well as the observational constraints.

\subsection{Redshift Distributions}\label{subsec:redshift_distributions}

In Figure~\ref{fig:miri_counts_redshift_split} we examine the relative contribution of sources in different redshift regimes. The dominant contribution for $S_{\nu}>0.1$\,$\mu$Jy sources comes from galaxies at \(z \lesssim 2\). 

\begin{figure*}
    \centering
    \includegraphics[width=\textwidth]{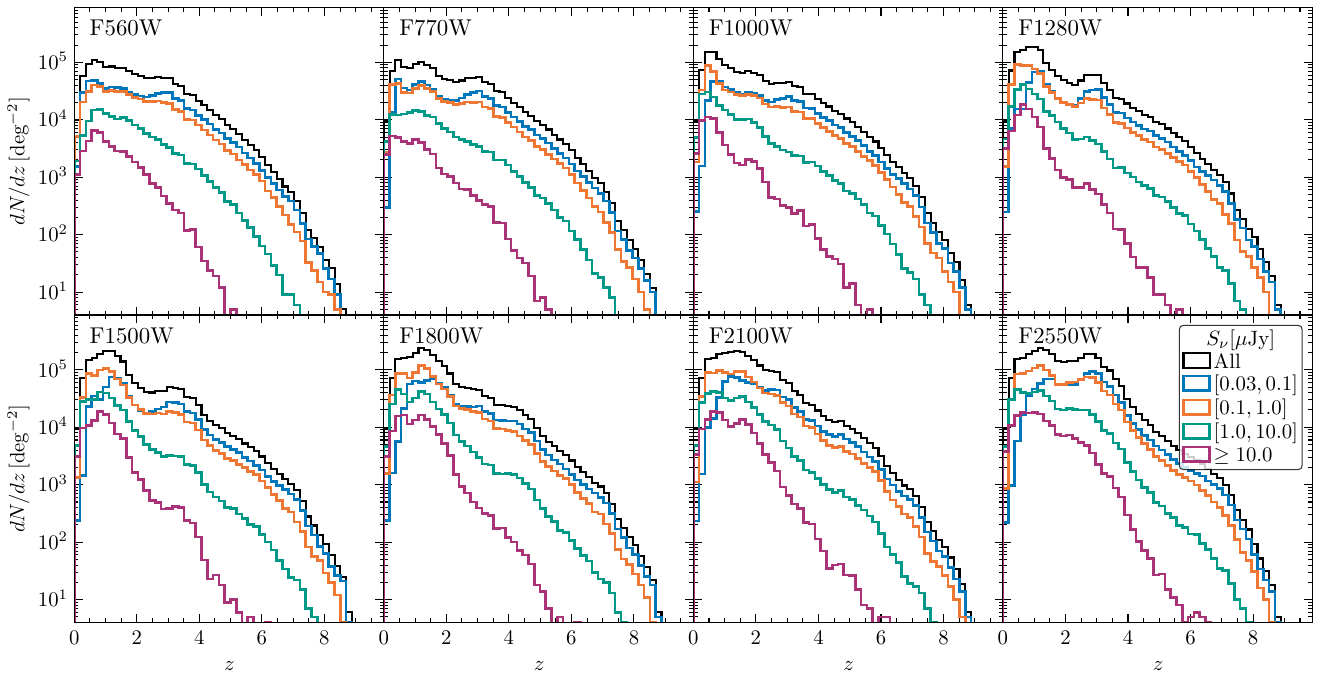}
    \caption{The redshift distributions across 8 MIRI imaging filters for the modified pySIDES model with \(\alpha = 0.55\).  The distributions are separated by cumulative contribution with varying detection thresholds starting with a flux limit of \(S_\nu > 0.03 \, \mu\rm Jy\).}
    \label{fig:redshift_distributions_by_flux_bins}
\end{figure*}

In Figure~\ref{fig:redshift_distributions_by_flux_bins}, we examine the redshift distributions in different observed flux density ranges. We often observe a bi-modal or even more complex redshift distribution. This arises from major features in the mid-IR SEDs coming into different filters. For example at the shorter wavelengths (F560W, F770W, and even F1000W) the counts are boosted at \(z \sim 3\) by the negative $k$-correction as the redshifted SED samples up towards the \(1.6 \, \mu \rm m\) stellar bump. For the longer wavelength filters, the multi-peak behavior arises as the PAH features \(3.3, 6.2, 7.7, 8.6, 11.3 \, \mu \rm m\) come in and out of the different filters. As a result, MIRI surveys that reach to 0.1-0.2\,$\mu$Jy (similar to H2P and SMILES) probe significant numbers of galaxies at cosmic noon and even beyond, although the counts drop significantly at $z>3$ for these flux levels.  

\begin{figure*}
    \centering
    \includegraphics[width=\textwidth]{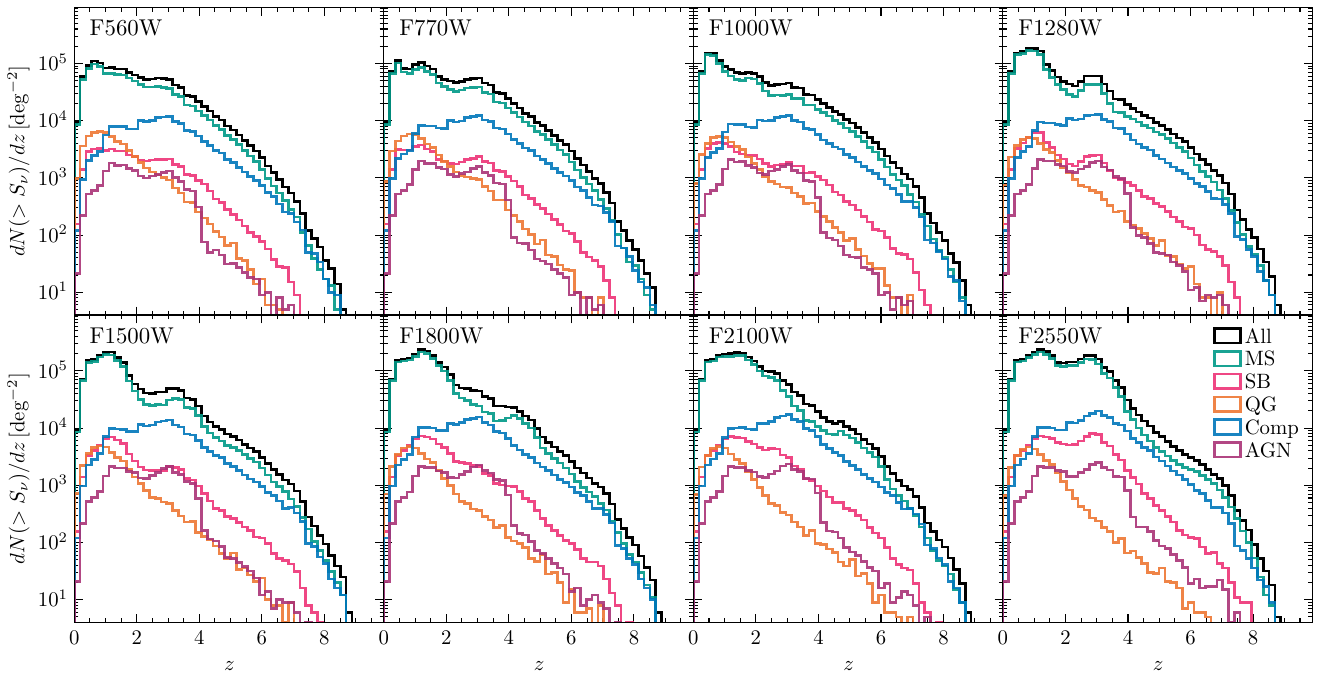}
    \caption{The redshift distributions across 8 MIRI imaging filters for the modified pySIDES model with \(\alpha = 0.55\). The distributions are separated by cumulative contribution from each population: main Sequence (MS), starburst (SB), quiescent (QG), composite (Comp), and AGN. }
    \label{fig:redshift_distributions_by_population}
\end{figure*}

Lastly, in Figure~\ref{fig:redshift_distributions_by_population} we show the redshift distributions for galaxies with \(S_\nu > 0.03\,\mu \rm Jy\), split into main sequence (MS), starburst (SB), quiescent (QG), composite galaxies (Comp), and AGN. The sample is overwhelmingly dominated by main-sequence galaxies, consistent with the F560W observations from \citep{Sajkov2024}. This is also consistent with the predictions of \citet{Cowley2018}, who find that MIRI-selected samples are dominated by disk-mode star formation. Quiescent galaxies contribute primarily at low redshift \(z \lesssim 1\), but do not dominate the population at any redshift.

Composite galaxies (\(f_{\rm AGN, MIR} \in [0.2-0.8]\)) represent the second most abundant population across all redshifts and wavelengths, especially significant among the cosmic noon ($z\sim1-3$) MIRI sources. AGN and Starburst galaxies exhibit similar redshift distributions with primary peaks at \(z \sim 1\) and a secondary peak around cosmic noon. The quiescent population declines sharply past \(z \gtrsim 1\) in line with expectations of cosmic evolution as well as due to the lack of strong mid-IR features to boost their observed emission in the MIRI filters.

\begin{figure}
    \centering
    \includegraphics[width=\columnwidth]{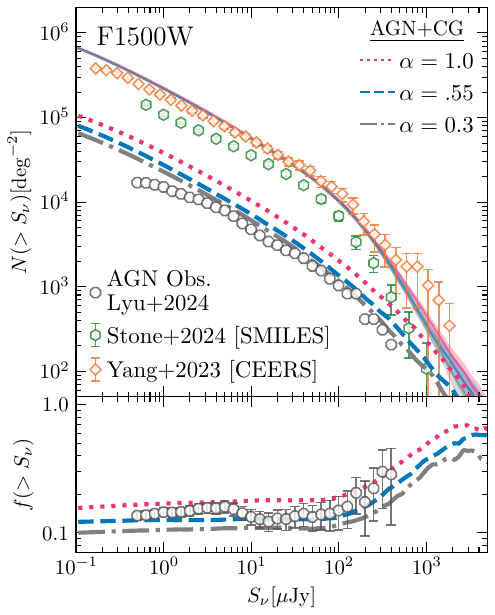}
    \caption{The effect of different \(\alpha\) values on the modified pySIDES model for the MIRI F1500W filter. Upper panel: Integral counts for \(\alpha = 0.3, 0.55, 1\). The total counts are shown as thin solid lines, while the AGN and Composite combined (AGN+Comp) contributions are shown with dotted, dashed, and dash-dotted lines, respectively. Observed counts from \citep[][]{Stone2024, Yang2023} are shown for comparison with the total counts. Bottom panel: AGN+Composite fractions as a function of observed flux. Both: Observed AGN (including SF-AGN mixed sources) from \citet{Lyu2024} are overplotted for comparison.}
    \label{fig:alpha_dependence}
\end{figure}

\subsection{The contribution of AGN and starburst-AGN composites to the MIRI counts}\label{subsec:counts_by_population}
\begin{figure*}
    \centering
    \includegraphics[width=\textwidth]{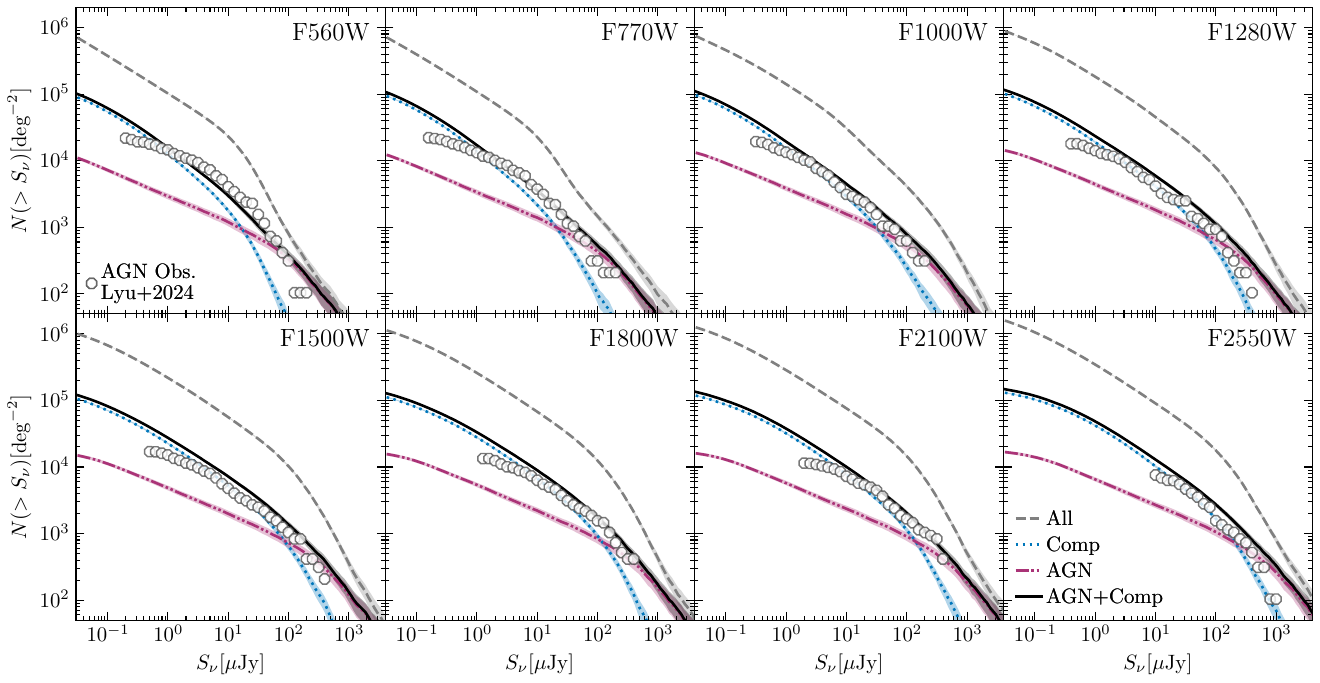}
    \caption{The modified pySIDES integral number counts across eight MIRI imaging filters assuming \(\alpha = 0.55\). The composite (Comp), AGN, and their combined contributions are shown in blue, purple, and black, respectively. Observational AGN (including SF-AGN mixed) counts from \citep{Lyu2024} are shown as gray markers for comparison with predictions. The total integral counts (gray) and data points from various observational programs (see Section \ref{subsec:number_counts} for details) are also shown for comparison. The relatively poor match at the faint end of the F560W and F770W bands is due to the conservative selection adopted when the MIRI data are limited in \citet{Lyu2024}.}
    \label{fig:miri_counts_population_split}
\end{figure*}

A key motivation for our updated pySIDES model has been to include explicitly the contribution of AGN and starburst-AGN composites. These, however, are subject to model uncertainties such as the value of the $\alpha$ parameter from Equation~\ref{eq:disk_and_agn_luminosity}. This parameter varies for Type I, II, and for obscured AGN \cite[e.g.][]{Lyu2017}. To assess its implication, we derive $f_{\rm AGN}$ ECDFs (see Section \ref{subsec:determining_f_agn}) using \(\alpha = 0.3, 0.55, 1\). Figure~\ref{fig:alpha_dependence} shows the AGN number counts for these three $\alpha$ models. We find a negligible dependence on \(\alpha\) in the total counts as shown in the upper panel. The difference becomes apparent however when looking at the AGN and Composites+AGN integral counts. The bottom panel shows the fraction of all galaxies that are composites+AGN as a function of flux density. Compared with the data from \citet{Lyu2024}, we find the best agreement with $\alpha=0.55$. This is somewhat expected because the normal AGN template with \(\alpha \approx 0.55\) reproduces most ($>70\%$) of the AGN IR observations on a statistical basis \citep{Lyu2022}. The same template, with attenuation added at shorter wavelengths, was adopted in  \citet{Lyu2024} and matches the relevant {\sl JWST} photometry. We therefore select the \(\alpha = 0.55\) model for exploring the contribution of AGN and composites across all MIRI filters. 

In Figure~\ref{fig:miri_counts_population_split}, the number counts of AGN,  composite galaxies (Comp), and combined composites+AGN (see Section~\ref{subsec:determining_f_agn} for definitions) are shown alongside the total counts as well as the observed AGN counts from \citet{Lyu2024}. Our model predicts that strong AGN dominate the bright end of the counts while composites contribute more at intermediate-to-faint fluxes. The AGN contribution is highest at the highest mid-IR fluxes, consistent with AGN being more common among higher luminosity sources \citep[see e.g.][]{Kirkpatrick2015}. 

As noted by \citet{Lyu2024}, AGN counts are conservative estimates and can vary by up to a factor of two depending on the selection criteria. An additional complication arises from low-z dwarf galaxy populations, whose star-heated dust SED may have some evolution, making AGN identification among these systems challenging. Taking this into consideration, the combined AGN and composite counts agree with the observed AGN number counts. At the longer wavelengths, the reported AGN counts are limited by the relatively shallow depth of existing surveys, a limitation that future deep MIRI observations are expected to overcome. We further explore the role of cosmic variance in shaping the bright end of the AGN counts in Section~\ref{subsec:cosmic_Variance}.

\subsection{The relative fractions of AGN and starburst-AGN composites}\label{subsec:population_fractions}

In Figure~\ref{fig:redshift_distributions_by_population} we already saw that $z<3$ main sequence star-forming galaxies comprise the bulk of the MIRI sources in our model with starburst and quiescent galaxies forming much smaller fractions of the total. This is consistent with the findings of \citet{Sajkov2024} for the F560W population. Here we are examine in particular in the relative contribution of AGN and starburst-AGN composites as a function of flux density as compared with observational constraints from \citet{Lyu2024}. 

\begin{figure*}
    \centering
    \includegraphics[width=\textwidth]{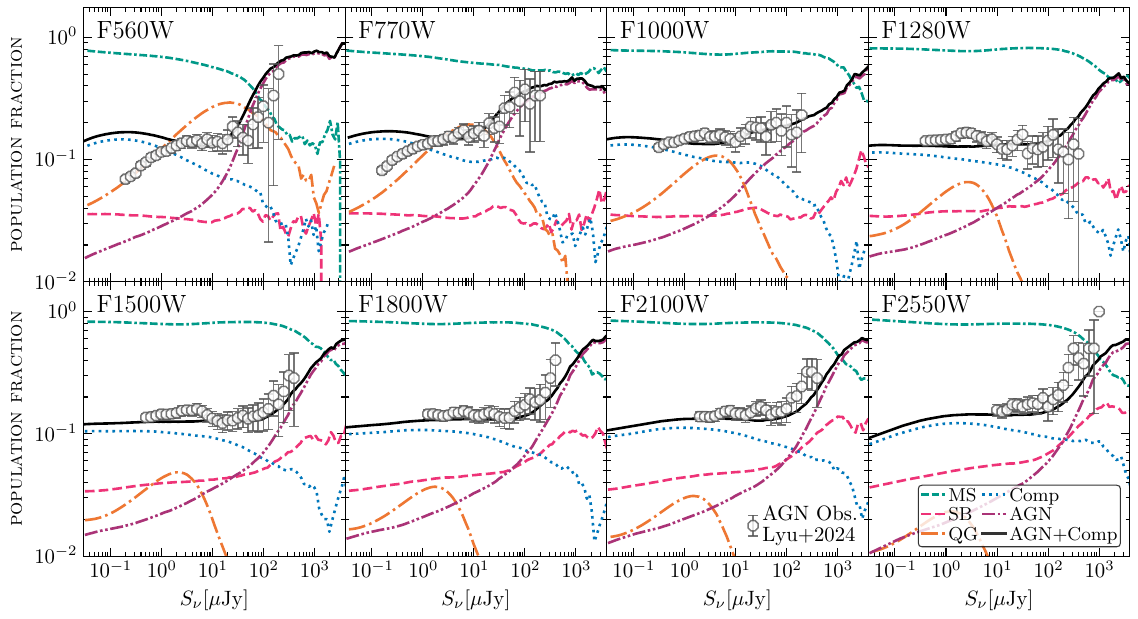}
    \caption{The cumulative MIRI flux distribution from the modified pySIDES model for Main Sequence (MS), Starburst (SB), Quiescent (QG), Composite Galaxies (Comp), AGN, and AGN+Composite Galaxies. Fractional contributions were computed by normalizing the cumulative counts of each population to the total counts at a given flux. Observed AGN (including SF-AGN mixed) fractions from \citet{Lyu2024} are overplotted as gray markers for comparison.}
    \label{fig:miri_counts_population_fraction}
\end{figure*}

Figure~\ref{fig:miri_counts_population_fraction} shows relative fractions of Main Sequence (MS), Starburst (SB), Quiescent (QG), Composite (Comp), and AGN. We overlay the AGN fractions from \citet{Lyu2024}. As discussed above, the AGN dominate the bright end, consistent also with the MIPS \(24 \, \mu \)m results from \citep{Kirkpatrick2015} at \(S_{24} >10^2 \, \mu \rm Jy\). When comparing with \citet{Lyu2024} for our AGN population only, we find good agreement at the bright end; however, our modeled AGN population falls below the observation at \( \lesssim 10^2 \, \mu \rm Jy\). Including composite galaxies as well leads to a much closer match in the intermediate flux density regime where the AGN only model diverges from observations. When considering the combined (AGN+Comp, solid black) contributions, our modeled fractions are fully consistent with the observations.

Lastly, \citet{Lyu2024} stress that the reported number counts of AGN are sensitive to selection criteria and their AGN numbers may therefore represent a lower limit. We find that, using $\alpha = 0.55$  the composite+AGN combined fractions are within the observed scatter except for the lowest flux densities at F560W and F770W where our model exceeds the observed AGN fractions. This may be a regime where composites are under-counted in the current AGN selection. At higher flux levels, while the overall numbers are consistent, the shapes of the observed AGN fractions are at times different from those of our models -- see for example the AGN+composites fraction at F560W at $\sim$10\,$\mu$m. In all such cases there is a corresponding peak in the modeled quiescent fractions. It is possible that there is still some discrepancy where either some true quiescent galaxies are incorrectly assigned as observed AGN or vice versa in the models. 

\subsection{Effect of Cosmic Variance}\label{subsec:cosmic_Variance}
The shaded confidence intervals shown in Figures\,\ref{fig:miri_counts}-\ref{fig:miri_counts_population_split} only account for the statistical (Poisson) uncertainties and do not capture the effect of cosmic variance. For small survey fields like those with MIRI, cosmic variance is a significant issue \citep{Trenti_and_Stiavelli2008} since a field can fall within a relatively overdense or underdense region of the cosmic web, leading to variations in the measured counts \citep{Sajkov2024}. These effects are more pronounced for bright, rare sources, but their impact diminishes when combining observations from multiple independent fields or analyzing fainter populations \citep[e.g.][]{Moster2011}.

We estimate the impact of cosmic variance by comparing the predicted redshift distributions of sources with $S_{F560W}>0.1$\,$\mu$Jy with the observations from \citet{Sajkov2024}. In that study, only sources in two of their 8 pointings \((A_{\rm eff} \approx 4.6\, \rm acrmin^2)\) were cross-matched with COSMOS2020 \citep{Weaver2022}, which provided them with photometric redshifts estimates.

Rather than uniformly down-sampling our \(2\,\rm deg^2\) light cone to match the total area,  which would incorrectly assume a homogeneous source distribution, we model the observational strategy by simulating composite survey fields made from multiple MIRI-like pointings. Each individual pointing is a square aperture with \(A_{\rm eff} = 2.3 \, \rm arcmin^2\), approximating the MIRI field-of-view (FOV). We need two pointings to compare with the observations from \citet{Sajkov2024}. To avoid edge related artifacts, we ensure that each sampled field lies entirely within the RA and DEC boundaries of the simulated light cone.

\begin{figure}
    \centering
    \includegraphics[width=\columnwidth]{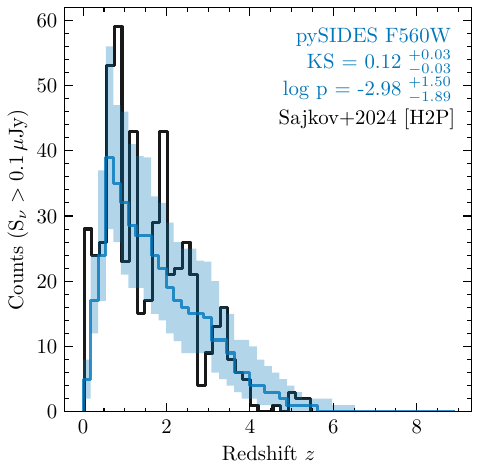}
    \caption{Redshift distributions of our modified pySIDES at F560W compared to cross matched sources in COSMOS2020 as presented in \citet{Sajkov2024}. The shaded blue regions represent the \(1 \sigma\) confidence interval when performing our Monte Carlo sub-sampling technique described in Section~\ref{subsec:cosmic_Variance}. We use 2 pointings with \(A_{\rm eff} = 4.6 \,\rm arcmin^2\) and 5000 draws to illustrate the effect of field-to-field variations. We perform a KS test to demonstrate that our model is statistically significant compared to the observed data distribution.}
    \label{fig:F560W_redshift_distribution}
\end{figure}

We repeat this process 5000 times, computing the median and 1\(\sigma\) confidence interval of the redshift distribution across the samples. Fig~\ref{fig:F560W_redshift_distribution} shows the resultant median and 1\,$\sigma$ spread relative to the measured redshift distribution from \citet{Sajkov2024}. For each monte-carlo realization, we perform a two-sample Kolmogorov-Smirnov (KS) test against the observed distribution. The two distributions are qualitatively similar, and the median and 1\(\sigma\) spread show good agreement. There is a spread in the KS $p$ values due to cosmic variance, but they suggest the simulated and observed redshift distributions are statistically distinct. Some of the differences can be attributed by 15\% of the MIRI F560W sources in \citet{Sajkov2024} not having matches in COSMOS2020. These sources would most likely boost the $z\sim1-5$ portion of the observed redshift distribution, leading to closer agreement with the model. We are also affected by small number statistics where each count per bin in the observed redshift distribution has a non-negligible Poisson error. 

Note that our cosmic variance modeling is simplified. For example, the MIRI field is an elongated rectangle rather than a square, and we ignore the Lyot region with \(25^"\times 25^"\) FOV. But these are secondary effects compared to the difference in scale between the \(4.6\,\rm arcmin^2\) survey vs. our \(2\,\rm deg^2\) lightcone. 

The effects of cosmic variance are particularly strong for the rarer bright AGN within the MIRI-selected samples. To quantify this, we extend the approach described above and measure how the total AGN counts fluctuate as a function of simulated survey area. For each target survey area, we tile \(N\) independent pointings (\(A_{\rm eff} = N \cdot 2.3 \, \rm arcmin^2\)) until the desired survey area is reached. We generate 5000 realizations per survey area. This allows us to calculate the error (i.e. the standard deviation among the 5000 realizations) on the AGN counts for a given flux limit. Since brighter flux limits yield fewer AGN, comparing the raw variance can be misleading. Therefore, we normalize the simulated error by the mean total AGN counts \(\bar N(>S)_{\rm AGN}\, [\rm arcmin^{-2}]\), and report the dimensionless variance (\(\sigma_{\rm AGN}/\bar N(>S)_{\rm AGN}\)). 

\begin{figure}
    \centering
    \includegraphics[width=\columnwidth]{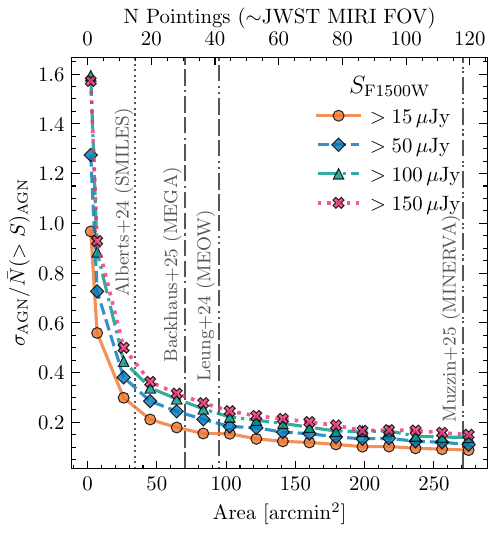}
\caption{The fractional error of total AGN counts as a function of increasing simulated survey area. The lower x-axis label represents the effective survey area, while the upper x-axis label indicates the corresponding number of independent MIRI-like pointings. Different flux thresholds in the F1500W band are overlaid to show how cosmic variance evolves with area and source brightness as described in Section~\ref {subsec:cosmic_Variance}. The area coverage of SMILES, MEGA, MEOW, and MINERVA (expected) is indicated by vertical lines.}
    \label{fig:cosmic_variance_over_rho}
\end{figure}

Figure~\ref{fig:cosmic_variance_over_rho} shows how this quantity evolves with increasing survey area for several flux thresholds in the F1500W band. We find that cosmic variance decreases with increasing area as expected. The scatter is particularly large with a slower convergence at brighter flux limits, where AGN are rarer and more clustered. Because our method assumes fully independent pointings, it would underestimate the true error for surveys of the same area but with tiled pointings. Figure~\ref{fig:cosmic_variance_over_rho} shows that for studies of mid-IR AGN, surveys with areas below about 20-25\,arcmin$^2$ will suffer greatly from cosmic variance. 

For comparison, the SMILES MIRI survey used by \citet{Lyu2024} covers \(\sim 34\, \rm arcmin^2\), which lies in the regime where the error is flatter. However, even here the fractional error due to cosmic variance is still substantial; for a survey of roughly the area of SMILES the fractional error is on the order of 30\% at the bright end (e.g. $S_\nu>50$\,$\mu$Jy). Other multi-band MIRI imaging surveys such as MEGA \citep{Backhaus2025} and MEOW \citep{Leung2024, Teodora2026} cover \(\sim 70\,\rm arcmin^2\) and \(\sim 95\,\rm arcmin^2\)  respectively, placing them near the transition regime where the fractional uncertainty due to cosmic variance drops below \(\sim 20\%\). To reduce the impact of cosmic variance to \(\lesssim 20\%\) across a wide flux range, we suggest future MIRI surveys target at least \( \sim 60-80 \rm \, arcmin^2\). For a truly cosmic-variance–limited measurement ($\lesssim$10–15\% uncertainty), we recommend \(\gtrsim 125 \rm \, arcmin^2 \) (\(\gtrsim\) 50 independent MIRI pointings), ideally distributed across multiple widely separated fields. 

An ongoing joint NIRCam medium-band survey with complementary MIRI imaging (F1280W and F1500W) is addressing this issue by covering \(\sim 275 \rm \, arcmin^2\) across multiple well-studied legacy fields \citep[MINERVA][]{Muzzin2025}. In Figure~\ref{fig:cosmic_variance_over_rho}, vertical lines indicate the F1500W survey areas of SMILES (34.5 arcmin\(^2\)), MEGA (70 arcmin\(^2\)), and MINERVA (275 arcmin\(^2\)), highlighting how increasing survey area and a multi-field design (in the case of MINERVA) mitigates the effects of cosmic variance.

\section{Discussion}\label{sec:discussion}
\subsection{Population differences between models}\label{subsec:population_differences_modeling}
\begin{figure}
    \centering
    \includegraphics[width=\columnwidth]{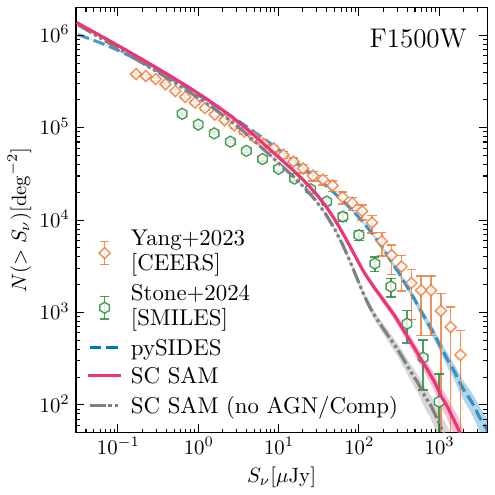}
    \caption{The integral number counts in the \(\,S_{\rm F1500W} > 0.68 \,\mu \rm Jy \) (e.g. SMILES \(5\sigma \) sensitivity limit \citep{Alberts2024}). We show modified SC SAM using the updated SED library and without using the updated SED library (gray). Here we adopt \(\alpha = 1\).}
    \label{fig:sc_sam_no_agn}
\end{figure}

The SC SAM model was previously compared against the CEERS counts in \citet{Yang2023_imaging}, where it was found to underestimate the counts at the bright end and overestimate them at the faint end. Although our modified pySIDES and SC SAM lightcones both adopt the same IR library, pySIDES reproduces the observed counts better. Nevertheless, our approach shows significant improvement to the SC SAM tested in \citet{Yang2023}, including better agreement at the faint and bright of the observed counts. This is due to improved dust templates as well as the addition of AGN and composite templates. Figure~\ref{fig:sc_sam_no_agn} shows that including AGN and composite populations brings the SC SAM predictions closer to the observations, although they remain systematically lower than both the observed counts and the pySIDES predictions. This is true even at the most extreme ($\alpha=1$) case. Because both the pySIDES and SC SAM models rely on the same IR SED templates, difference arises from variations in the underlying galaxy populations predicted by the models. 

\begin{figure*}
    \centering
    \includegraphics[width=\textwidth]{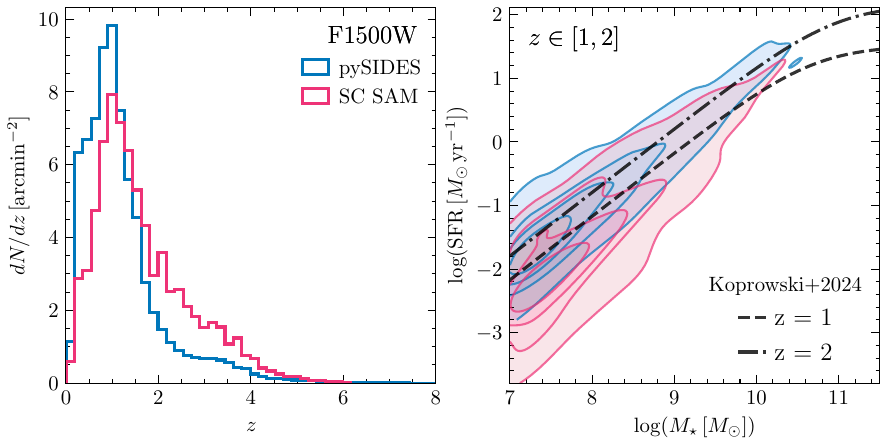}
    \caption{Left: The redshift distribution in the F1500W using the \(5 \sigma = 0.68 \, \mu \rm Jy \) sensitivity limit from SMILES \citep[][]{Alberts2024}
    for the modified pySIDES and SC SAM. Right: The SFR vs stellar mass in the redshift range \(z \in [1,2]\). We overlay the observational constraints from \citet{Koprowski2024} for comparison. }
    \label{fig:sides_SC_SAM_population_comparison}
\end{figure*}

In Figure~\ref{fig:sides_SC_SAM_population_comparison}, we explore the differences in the underlying population for detected sources in F1500W with  \(\,S_{\rm F1500W} > 0.68 \,\mu \rm Jy \) \citep[e.g. SMILES \(5\sigma \) sensitivity limit;][]{Alberts2024}. The left-hand panel shows the F1500W redshift distributions. The modified pySIDES model shows a peak at \(z \sim 1\) and drops off past \(z \gtrsim 2\). The SC SAM display a bimodality at \(z \sim 1 \,\mathrm{and}\, z \sim 2\). The SC SAM sources are at somewhat higher redshifts on average, making them faint and consistent with Figure~\ref{fig:sc_sam_no_agn}, where the counts are slightly overestimated at the faint end. The right-hand panel of Figure~\ref{fig:sides_SC_SAM_population_comparison} shows the SFR vs stellar mass at $z=1-2$, excluding quiescent galaxies. We find that both pySIDES and SC SAM are consistent with the star-forming main sequence reported by \citet{Koprowski2024}, despite the SC SAM galaxies being biased toward $z\sim2$ and the pySIDES galaxies being more biased toward $z\sim1$. Therefore, we conclude that the main factor leading to the discrepancy in MIRI count predictions between the two models is that galaxies build up faster in the SC SAM models than in the pySIDES one.

\subsection{Application of models to NIRCam+MIRI surveys}\label{subsec:applications}

In this paper, we have focused on modeling the MIRI source population. However, what is fundamentally needed is a fully consistent modeling framework that reproduces the observed properties of galaxies in both the optical (primarily stellar photospheric) and infrared (dust emission) parts of the spectrum. This is particularly relevant given upcoming {\sl JWST} surveys with significant overlap between the NIRCam and MIRI coverage \citep[e.g., MINEVRA;][]{Muzzin2025}. Being able to model the MIRI population with respect to the NIRCam one would be a particularly powerful tool. 

\begin{figure*}
    \centering
    \includegraphics[width=\textwidth]{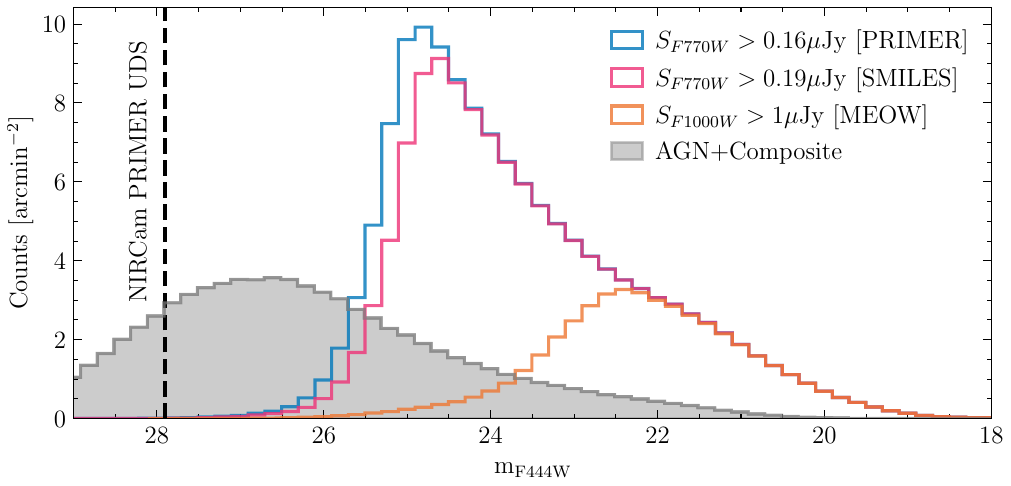}
    \caption{NIRCam F444W magnitude distributions from the modified SC SAM with $\alpha = 0.55$, shown for samples selected using MIRI flux-density limits corresponding to the SMILES, MEOW, and PRIMER surveys \citep{Stone2024, Leung2024, Donnellan2024}. The full AGN+Composite population is shown in gray. The vertical dashed line marks the PRIMER UDS $5\sigma$ depth at $m_{\rm F444W} = 27.9$.}
    \label{fig:NIRCam_vs_MIRI}
\end{figure*}

Figure~\ref{fig:NIRCam_vs_MIRI} illustrates how such a comprehensive modeling framework can be applied. We show the NIRCam F444W AB magnitude distribution of galaxies selected at different MIRI detection limits \citep[e.g., SMILES, MEOW, PRIMER;][]{Stone2024, Leung2024, Donnellan2024}. The vertical dashed line denotes the NIRCam F444W depth of the PRIMER Ultra Deep Survey (UDS) \citep{Donnellan2024}. The MIRI F777W-selected samples turn over at $m_{\rm F444W} \approx 25$, several magnitudes brighter than the NIRCam detection limit. This mismatch highlights the incompleteness of MIRI-selected samples relative to the full galaxy populations accessible with NIRCam. Given the large disparty between NIRCam and MIRI depths, we expect essentially all MIRI-detected sources to have NIRCam counterparts in regions of overlapping coverage. Multi-band NIRCam imaging is therefore critical for MIRI science, as it provides robust photometric redshifts and constraints on stellar populations. 

While the brightest AGN and composite systems are readily detected in MIRI, as also shown in Figure~\ref{fig:miri_counts_population_fraction}, our model predicts a substantial population of fainter AGN and composites that remains below current MIRI detection thresholds. The majority of these systems are composites in which stellar emission from the host galaxy dominates the optical and near-infrared light, making them difficult to distinguish from purely star-forming galaxies using NIRCam photometry alone. Accessing this population will likely require complementary diagnostics from upcoming optical \citep[e.g.,][]{Greene2022} and far-infrared spectroscopic surveys \citep[e.g.,][]{Rodgers2023}.

\begin{figure*}
    \centering
    \includegraphics[width=\textwidth]{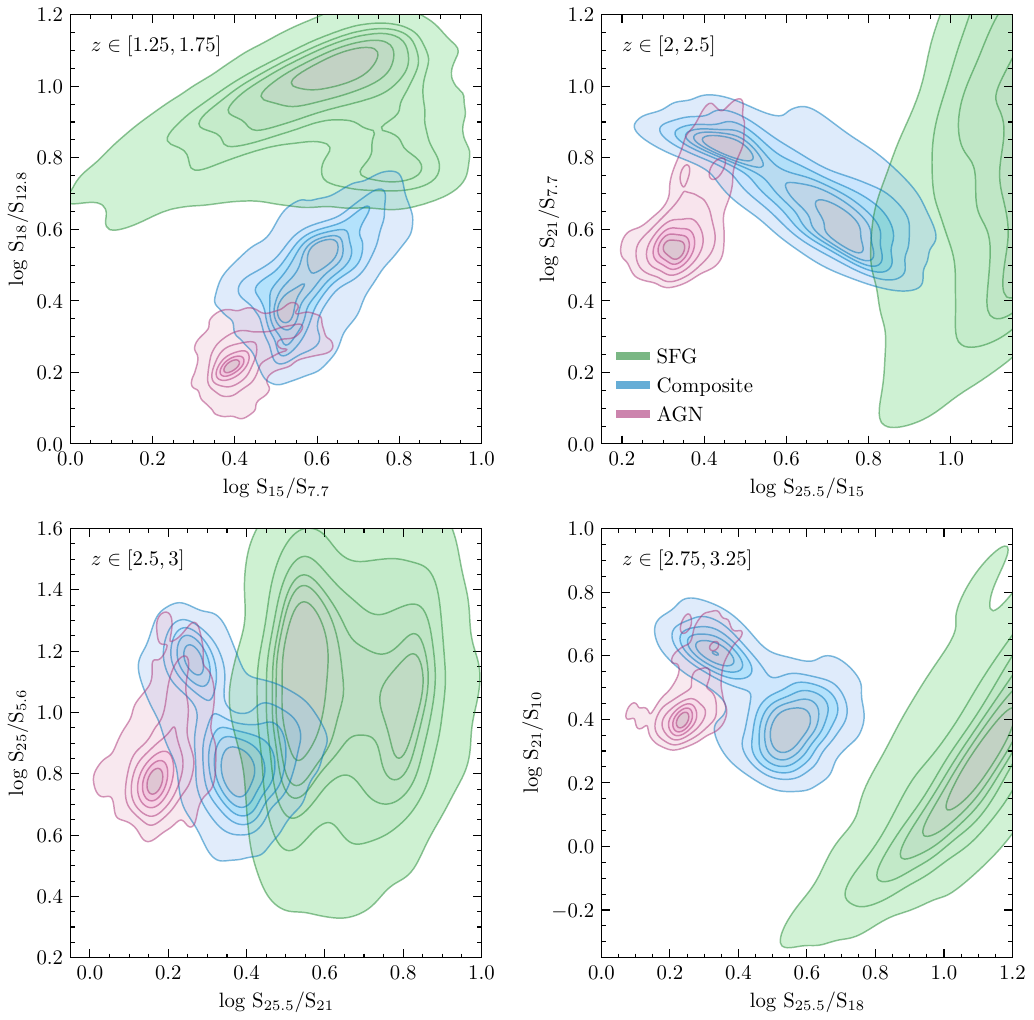}
    \caption{MIRI color-color diagnostic diagrams separating Star Forming galaxies (SFG), composite galaxies, and AGN, from the modified pySIDES model, see also \citet{Kirkpatrick2017}. Photometric cuts and noise estimates are matched using the reported \(5 \sigma\) sensitivity limits from the SMILES survey \citep{Alberts2024}. The KDE contours in this figure are made available.}
    \label{fig:color_space_diagrams}
\end{figure*}

Lastly, in Figure\,\ref{fig:color_space_diagrams} we show four observed color--color combinations that were found to best separate the different populations in four redshift slices from $z\approx1.5$ to $z\sim3$. This figure is an extension of the diagnostic diagrams already presented in \citet{Kirkpatrick2017}, however, it extends to higher redshift \(z \simeq 3\). We also adopt the observed depths from the SMILES survey \citep[][]{Stone2024} by both cutting out sources with $<5\sigma$ in each band as well as including Gaussian noise of the corresponding $\sigma$ per filter. These depths and the explored filter combinations are particularly relevant for the ongoing MINERVA survey \citep{Muzzin2025}, which will reach comparable depths over the largest multi-band MIRI area (275 arcmin\(^2\)) in F1280W and F1500W. We also explored these color-color diagrams at shallower flux limits and find that while the centers of the distributions remain unchanged, the contours broaden slightly due to increased scatter. The Kernel Density Estimation (KDE) contours shown in this figure are provided in a machine-readable format to facilitate future analyses.

\subsection{Comparison with literature}\label{subsec:compare_lit_future_work}
Here we consider how our modeling framework to predict the MIRI number counts and especially the contribution of AGN therein compares with other approaches in the literature. Below we specifically consider the semi-analytic models of \citet{Cowley2018} and \citet{Lagos2019} as well as the empirical models of \citet{Bisigello2021} and \citet{Rowan-Robinson2025}.

\citet{Cowley2018} uses the GALFORM semi-analytic model coupled with GRASIL \citep{Silva1998} to generate the IR SEDs. While GALFORM incorporates AGN feedback when modeling the galaxies star-formation histories, there are no explicit AGN or AGN-composites, which means the model does not make specific predictions for the contribution of AGN. GALFORM predicts that galaxies forming stars in the disk,\footnote{In GALFORM, the ``quiescent'' classification refers to steady, disk-mode star formation and is therefore most directly comparable to observationally defined main-sequence galaxies.} dominate the source counts across all MIRI filters, in agreement with both our results and observational constraints from \citet{Sajkov2024}. In Figure~\ref{fig:miri_counts}, we include the count predictions from \citet{Cowley2018} and find that they tend to overpredict the faint-end counts at shorter wavelengths (e.g., F560W and F770W), indicating a higher predicted abundance of disk-mode star-forming galaxies relative to observations.
 In this figure, we use their model with evolving feedback rather than the baseline model since the former is closer to the observed counts. 

\citet{Lagos2019} models the MIRI counts using the SHARK semi-analytic model. This model offers a physically motivated framework to model the IR galaxy populations and adopts the \citet{Dale2014} IR SED templates to model dust emission. This SED library includes both star forming galaxies and AGN templates (classified by their 5–20 \(\mu\)m mid-IR strength) with the mid-IR portion constructed from the average {\sl Spitzer}-based observed SEDs. SHARK v1.0 assumed minimal AGN emission, yet they achieve good agreement with the \(7.7\, \mu\)m number counts \citep[e.g see;][]{Stone2024, Harish2025} and reproduce galaxy populations from far UV to FIR \citep{Lagos2019}. The updated SHARK v2.0 \citep[][]{Lagos2024, Bravo2025} now includes the AGN modeling across cosmic time, and predictions of AGN counts from this framework would provide a valuable comparison to the models presented here.

\citet{Bisigello2021} uses the empirical Spritz model where galaxy populations are built starting with the observed {\sl Herschel} IR luminosity functions. Mock catalogs are then constructed for various galaxy populations based on empirical scaling relations. This model was compared with the SMILES F560W-F1000W counts in \citet{Stone2024}, yielding excellent results. We cannot find explicit mid-IR AGN counts comparisons with the literature for Spritz. Spritz generates images and associated diagnostics by constructing lightcones that reproduce the observed two-point correlation function, whose normalization varies with stellar mass. This is contrary to our work, which relies on dark matter simulations and halo merger histories to generate mock images \citep[e.g.][]{Bethermin2017, Bethermin2022, Yung2022_VI}

On the other hand, Spritz also incorporates nebular emission arising from both star formation and AGN into the IR. This also allows them to make predictions for the future far-IR spectroscopic PRIMA mission in \citet{Bisigello2024}. As we do here, \citet{Bisigello2024} compares the Spritz results with the SC SAM and found that only 4\% of galaxies in SC SAM have a BHAR \(>0.06 \, M_\odot/ \rm yr\) compared to 44\% in Spritz. This would lead to a much lower number of AGN and composite systems. This observation would also explain the underestimation of the bright-end counts from the SC SAM that we also find (see Fig.~\ref{fig:sc_sam_no_agn}).  

Lastly, \citet{Rowan-Robinson2024} and \citet{Rowan-Robinson2025} present an empirical model where they start with the evolving 60\,$\mu$m luminosity function and consider a mix of five different populations, including AGN and quiescent galaxies.  There is no explicit modeling for composite systems. The SEDs templates adopted for the different populations are observationally driven and often use specific galaxies as prototypes (e.g. M82 and Arp220 for the starbursts). In Figure\,\ref{fig:miri_counts} we include the counts predictions from \citet{Rowan-Robinson2025}. This model shows excellent agreement across the shorter F560W and F770W. However, at longer wavelengths, the counts are overestimated in the flux range \(1-100 \,\mu\)Jy. 

\section{Summary \& Conclusions}\label{sec:Summary}
In this paper, we update the pySIDES  \citep{Bethermin2012,Bethermin2017, Bethermin2022} semi-empirical model to extend coverage to the observed frame 5-25\,$\mu$m regime covered by the {\sl JWST} MIRI filters. This was done by adding a stellar emission component to the pySIDES star-forming galaxies SED library as well as a quiescent galaxy SED template.  We also include an AGN SED library with AGN fractions ($f_{\rm AGN}$) ranging from 0 to 1. This library is a modified version of the AGN SED templates of \citet{Kirkpatrick2015}. Each galaxy is assigned a \(f_{\rm AGN}\) using $f_{\rm AGN}$ probability distributions in bins of \(L_{\rm IR,SF}\) and redshift that are constructed from the star-formation rates and black hole accretion rates that are part of the Santa Cruz SAM lightcone. We use these modifications to generate MIRI flux predictions for all galaxies in both the pySIDES lightcone as well as the SC SAM lightcone. Below we summarize our key results.

\begin{enumerate}
    \item Our updated pySIDES model reproduces well the observed number counts in the MIRI filters, including the shortest wavelength (F560W) dominated by stellar photospheric emission.  
    \item Unlike earlier models, we find that broadly cosmic noon, main sequence galaxies dominate the MIRI population. This is consistent with available observational constraints such as the population study of F560W MIRI sources in \citet{Sajkov2024}. 
    \item We are able to reproduce the counts of the MIRI-selected dusty AGN in \citet{Lyu2024}. We find that the brighter MIRI AGN (at $S_{\nu}\gtrsim20-50$\,$\mu$Jy) are likely dominated by their AGN, whereas at fainter fluxes are dominated by composite sources. 
    \item We investigate the effects of cosmic variance by randomly placing MIRI-like pointings (FOV \(\sim2.3\,\rm arcmin^2\)) within the \(2\,\rm deg^2\) lightcones. We are able to reproduce the observed roughly factor of 2 field-to-field variation seen in \citet{Sajkov2024} among individual MIRI pointings. We also quantify the fractional error \(\sigma_{\rm AGN}/ \bar N(>S)_{\rm AGN}\) for bright, rare AGN as a function of survey area. This suggests that surveys of order 30-40\,arcmin$^2$ such as SMILES have a roughly \(\sim 30\%\) cosmic variance error in the AGN counts. 
    
    \item To test the SC SAM framework with the same IR templates, we bypass the abundance matching and star formation property assignments, instead feeding it directly into the pySIDES procedure. We identify Main Sequence, Starburst, and Quiescent galaxies using their location on the sSFR vs M\(_\star\) diagram. We find that it underestimates the bright end of the counts likely as a result of the inherent population difference between pySIDES and SC SAM.

    \item We demonstrate how observers can use our lightcones by providing color–color diagrams that clearly distinguish AGN from star-forming galaxies. Thanks to the SC SAM’s inclusion of {\sl JWST} NIRCam photometry, we also examine the distribution of the F444W band at varying flux limits in F770W and F1000W. These flux thresholds are chosen to match the depths of existing and upcoming surveys such as SMILES, PRIMER, and MEOW \citep{Stone2024, Leung2024, Donnellan2024}
    
\end{enumerate}

We make the SED templates from Figure~\ref{fig:AGN_SED}, the number counts from Figure~\ref{fig:miri_counts}, the AGN and composite number counts from Figure~\ref{fig:miri_counts_population_split}, and the contours from Figure~\ref{fig:color_space_diagrams} publicly available. We also provide a Jupyter Notebook to load and interact with the data. All products are hosted on CANFAR \citep{Canfar2025} and are publicly available via a persistent DOI (\dataset[doi:10.11570/26.0001]{https://www.canfar.net/citation/landing?doi=26.0001}; \citealt{Vidal2026_data_products}).

\begin{acknowledgments}
We thank the anonymous referee for constructive comments that improved the clarity of this work. This work was supported by NASA Astrophysics Decadal Survey Precursor Science (ADSPS) grant 80NSSC25K0169. JL is supported in part by JWST Mid-Infrared Instrument (MIRI) grant No. 80NSSC18K0555, and the NIRCam science support contract NAS5-02105, both from NASA Goddard Space Flight Center to the University of Arizona. VU further acknowledges National Science Foundation (NSF) Astronomy and Astrophysics Research Grant 2536603 and STScI grant \#JWST-GO-01717.001-A, which was provided by NASA through a grant from the Space Telescope Science Institute, which is operated by the Association of Universities for Research in Astronomy, Inc., under NASA contract NAS 5-03127. 
\end{acknowledgments}





%
\facilities{The research reported in this paper was done on the Tufts University High Performance Computing Cluster \url{(https://it.tufts.edu/high-performance-computing)}.}

\software{Astropy \citep{Astropy2018}, numpy \citet{harris2020array}, \citep{2013A&A...558A..33A,2018AJ....156..123A,2022ApJ...935..167A},  pandas \citep{reback2020pandas}, matplotlib \citep{Hunter:2007}, Seaborn \citep{Waskom2021},}


\appendix
\renewcommand{\thefigure}{A.\arabic{figure}}
\setcounter{figure}{0}

Here we present two tests that we performed to validate our approach of including AGN in pySIDES.  In Section~\ref{sec:AGN_Luminosity_Functions}, we construct an AGN luminosity function from our pySIDES lightcone and compare it with literature constraints thereof. In Section~\ref{subsec:farir_counts}, we test the effects of our modified pySIDES on the predictions for the far-IR counts for which the original (no AGN) pySIDES is well known to match the observed counts very well \citep{Bethermin2012, Bethermin2017}. 

\section{AGN Luminosity Functions}\label{sec:AGN_Luminosity_Functions}
\begin{figure}
    \centering
    \includegraphics[scale = 1.2]{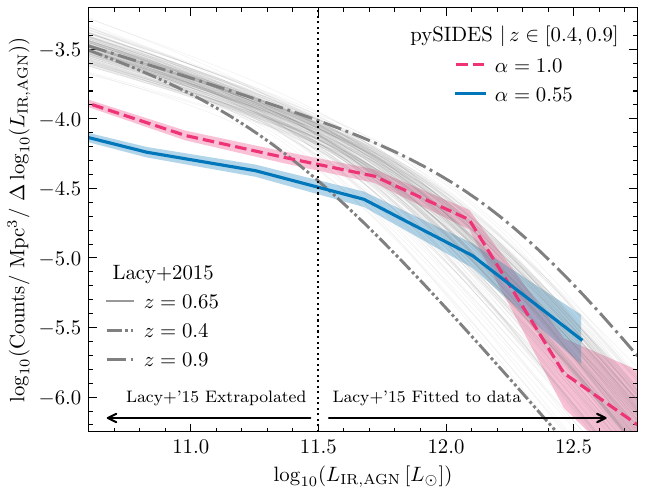}
    \caption{
    The $z = 0.4$–$0.9$ AGN luminosity function from the modified pySIDES model, including both AGN and composite galaxies ($f_{\rm AGN,MIR} \geq 0.2$) for \(\alpha = 0.55\) and \(\alpha = 1\). For comparison, we overlay observational constraints of the AGN luminosity function from \citet{Lacy2015}. The vertical dotted line indicates the lower bound of \(L_{\rm AGN}\) for which observational data were available. The dashed-dotted and dash-dot-dotted lines represent the best-fit functions from \citet{Lacy2015} \(z = 0.4\) and \(z = 0.9\), respectively. The thin lines represent the \(z = 0.65\) fits sampled from the normal distribution of the best-fit parameters and their reported uncertainties.}
    \phantomsection
    \label{fig:AGN_Luminosity_Function}
\end{figure}

We calculate the AGN luminosity function of the modified pySIDES model and include only sources classified as composite or AGN \(f_{\rm AGN,MIR} > 0.2\). The AGN infrared luminosity \(L_{\rm IR,AGN}\) is computed using Equation~\ref{eq:fraction_of_agn} and \ref{eq:total_luminosity_with_AGN}. To determine the comoving volume, we adopt the \citet{PlanckIII2016} cosmology as implemented in the Astropy \citep{Astropy2018} cosmology module. In Figure~\ref{fig:AGN_Luminosity_Function}, we show our computed AGN luminosity function from \(z \in [0.4,0.9]\) as well as observational constraints from \citep{Lacy2015}. The dashed-dotted and dashed-dot-dotted lines represent the best fit \(z=0.64\) and \(z=0.9\) luminosity function, assuming a fixed faint-end slope. The thin lines at \(z = 0.65\) are drawn by sampling from normal distributions centered on the best-fit parameters and their reported uncertainties. We conclude that within the \(L_{\rm IR,AGN}\) range where observational data are available, the model predictions lie within the reported scatter of observations.

\section{Far Infrared Counts}\label{subsec:farir_counts}
\begin{figure*}
    \centering
    \includegraphics[width=\textwidth]{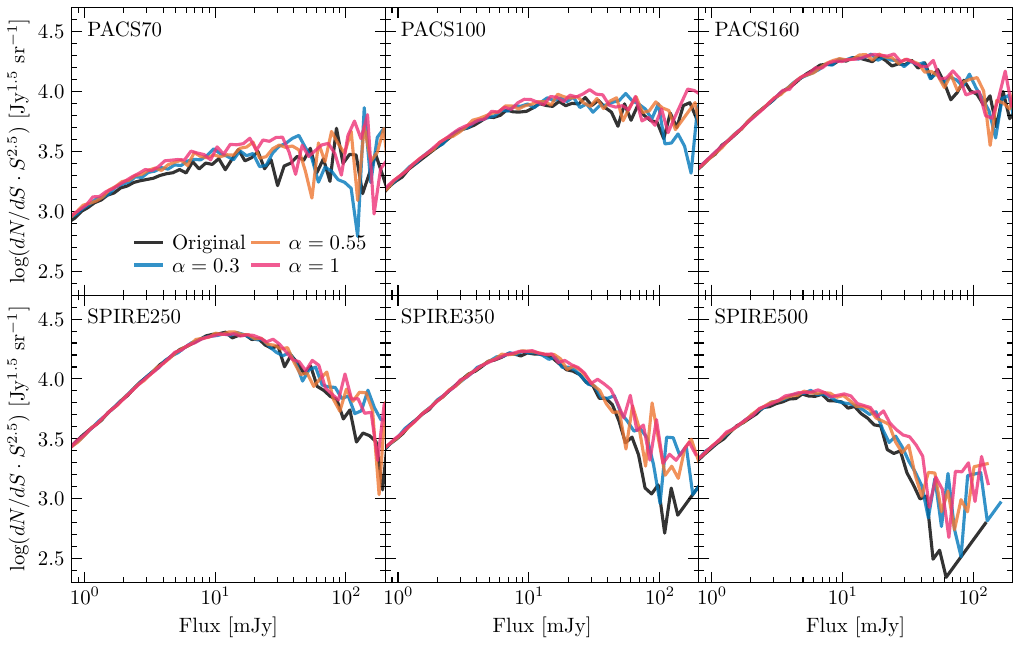}
    \caption{The Euclidean normalized differential number counts from {\sl Herschel} PACS/SPIRE compared the original implenations of SIDES \citep{Bethermin2017} and updated pySIDES models. The models with \(\alpha\) = 0.3, 0.5, and 1.0 are shown for comparison. 
}
    \phantomsection
    \label{fig:herschel_counts}
\end{figure*}

Figure~\ref{fig:herschel_counts} compares the Euclidean-normalized differential number counts from {\sl Herschel}-PACS at 70~\(\mu\)m to {\sl Herschel}-SPIRE at 500~\(\mu\)m for both the original and updated pySIDES models. The updated model closely reproduces the original results across the entire FIR wavelength range, demonstrating that our modifications do not compromise the model’s predictive coverage. We overlay our model predictions for \(\alpha = 0.3, 0.5,\) and \(1.0\). The influence of AGN emission is most pronounced at the shorter wavelengths (PACS 70~\(\mu\)m), however, this is within the relative scatter of the observed data points \citep[see Figure 4 in][]{Bethermin2017}. This comparison validates our updated pySIDES model as a reliable tool for investigating galaxy populations across the Mid-IR to Far-IR regimes.

\bibliography{bibliography}{}
\bibliographystyle{aasjournalv7}



\end{document}